\documentclass{article}
\usepackage[utf8]{inputenc}
\usepackage{amsmath,amsfonts,dsfont}
\usepackage{authblk}
\usepackage{amssymb}
\usepackage{tikz}
\usepackage[T1]{fontenc}
\usepackage{color}
\usepackage{latexsym}
\usepackage{amsbsy}
\usepackage{amstext}
\usepackage{graphicx}
\usepackage{amsmath}
\usepackage{amssymb}
\usepackage{tikz}
\usetikzlibrary{intersections, angles, quotes}
\usetikzlibrary{positioning}
\usepackage{tikz-cd}
\usetikzlibrary{arrows,decorations.markings}
\usetikzlibrary{quotes, calc}
\usepackage{mathtools}
\usepackage{amsthm}
\usepackage{amstext}
\usepackage{amssymb}
\usepackage{graphicx}
\usepackage{esint}
\usepackage{times}
\usepackage{color}
\usepackage{fancyhdr}
\usepackage{amsfonts,amsthm,amsmath,amssymb,graphicx,hyperref,youngtab}
\usepackage{amsmath,amsfonts,dsfont}
\usepackage{amssymb}
\usepackage{tikz}
\usepackage[T1]{fontenc}
\usepackage{latexsym}
\usepackage{amsbsy}
\usepackage{amstext}
\usepackage{graphicx}
\usepackage{amsmath}
\usepackage{amssymb}
\usepackage{tikz}
\usepackage{mathtools}
\usepackage{amsthm}
\usepackage{amstext}
\usepackage{amssymb}
\usepackage{graphicx}
\usepackage{esint}
\usepackage{times}
\usepackage{fancyhdr}
\def\C{\mathbb{C}}

\def\R{\mathbb{R}}

\def\H{\mathbb{H}}

\newcommand{\cA}{\mathcal A}

\newcommand{\cC}{\mathcal C}

\newcommand{\cJ}{\mathcal J}

\newcommand{\cL}{\mathcal L}
\newcommand{\cM}{\mathcal M}

\newcommand{\cZ}{\mathcal Z}

\newcommand{\be}{\begin{equation}}
\newcommand{\ee}{\end{equation}}
\newcommand{\bea}{\begin{eqnarray}}
\newcommand{\eea}{\end{eqnarray}}

\title{Thermodynamics of Isoradial Quivers and Hyperbolic 3-Manifolds}
\author{Ali Zahabi}
\affil{Institut de Math\'{e}matiques de Bourgogne, France}
\affil{National Institute for Theoretical Physics,
School of Physics and Mandelstam Institute for Theoretical Physics,
University of the Witwatersrand, SA}
\date{June 2017}
\date{}
\begin{document}
\maketitle
\tableofcontents
\begin{abstract}
{The BPS sector of $\mathcal{N}=2$, $4d$ toric quiver gauge theories, and its corresponding D6-D2-D0 branes on Calabi-Yau threefolds, have been previously studied using integrable lattice models such as the crystal melting model and the dimer model. The asymptotics of the BPS sector, in the large N limit, can be studied using the Mahler measure theory, \cite{Zah}. In this work, we consider the class of isoradial quivers and study their thermodynamical observables and phase structure. Building on our previous results, and using the relation between the Mahler measure and hyperbolic 3-manifolds, we propose a new approach in the asymptotic analysis of the isoradial quivers. As a result, we obtain the observables such as the BPS free energy, the BPS entropy density and growth rate of the isoradial quivers, as a function of the $R$-charges of the quiver and in terms of the hyperbolic volumes and the dilogarithm functions.
The phase structure of the isoradial quivers is studied via the analysis of the BPS entropy density at critical $R$-charges and universal results for the phase structure in this class are obtained.
Explicit results for the observables are obtained in some concrete examples of the isoradial quivers.}
\end{abstract}

\section{Introduction}
The toric quiver gauge theories lives in the world-volume of the D-branes on the tip of the Calabi-Yau threefolds $X$. The study of the quiver gauge theories motivated by string theory has been attracted a lot of interests after the seminal work, \cite{Do-Mo}. Among plausible approaches to explain the field contents, combinatorial and dynamical properties of the quiver gauge theories there is the dimer model and related toric geometry, that capture the properties of the toric quivers, \cite{Franco:2005rj,fr-Ma,Ha-Ke}.
The D6-D2-D0 bound states in $X$ are the BPS bound states of the toric quiver theory and they are determined by a closely related model to dimer model, namely the crystal melting model, \cite{Ok-Re-Va,Ya-Oo2}. In this work, we consider an interesting class of the dimer/crystal models, called isoradial models. We focus the quiver gauge theories associated with the isoradial dimer models and study the thermodynamical aspects of these quiver gauge theories using the techniques from the statistical mechanics of the isoradial models and related hyperbolic geometry. 

The properties of the BPS sector, including the BPS counting and its asymptotic behavior in the limit of large rank of the gauge groups and/or the large number of D0 branes are interesting nonperturbative problems. The asymptotic analysis of the BPS sector and the study of the thermodynamics and phase structure of the quivers contain interesting information about the gauge theories. Similar asymptotic problems in the chiral sector of the free quiver gauge theories have been studied recently, \cite{Ra-Wi-Za}.

In this work, we combine the tropical geometry methods introduced in the context of the toric dimer model \cite{Ke-Ok-Sh}, with the Mahler measure technology of the A-polynomials and related hyperbolic geometry. Using these methods, we explicitly compute the free energy and the entropy density of the isoradial quiver gauge theories in the asymptotic limit and extract the phase structure. The Mahler measure theory provides a powerful approach in the asymptotic analysis of many gauge theories and integrable systems. In our case, the Mahler measure of the Newton polynomial of the isoradial quiver and associated hyperbolic 3-manifolds determine the thermodynamical observables of the quiver. In our previous work \cite{Zah}, we introduced the connections between the Mahler measure and the BPS entropy density and growth rate of the quiver. Building on these results, in this work, we elaborate on the infinite class of the isoradial quivers and associated Mahler measure of the A-polynomials. 
The importance of the generalized Mahler measure techniques in the asymptotic analysis of the toric quivers are the computational techniques that they provide for explicit analysis of the asymptotics \cite{Zah}. In this work, we will use related computational techniques arising from the connections between the Mahler measure and the hyperbolic 3-manifolds. Moreover, this connection gives a geometric realization of the asymptotics of the toric quivers in the isoradial class.

One of our aim in this paper is to provide a geometric interpretation and expression for the BPS entropy. Our asymptotic analysis is based on the hyperbolic 3-manifold formulation of the thermodynamics of the isoradial quivers, their entropy density and the phase structure. We obtain new explicit mathematical computations for the Mahler measure of isoradial quiver, leading to new physical results for the thermodynamical observables in the context of the BPS sector of isoradial quivers, as a function of the $R$-charges of the quiver.

The main mathematical result of this work is the computation of the Mahler measure of the NP of isoradial quivers in terms of the hyperbolic 3-manifolds. Then, using our previous result in \cite{Zah}, we obtain our main physical results, i.e. explicit formulas,  as a function of the $R$-charges, for the observables of the isordial quiver such as entropy density. Another mathematical result of this work is the explicit computation of the Mahler measure of a generic example in the class of isoradial quiver, namely the cyclic $k$-polygon quiver, in terms of the Bloch-Wigner functions.

Having obtained the BPS entropy in terms of the hyperbolic volumes and dilogarithm functions, we perform further analysis to extract the phase structure of the BPS sector from the Bloch-Wigner and Lobachevsky functions. The phase structure of the isoradial quivers can be studied at critical limit of the entropy density. In this limit, the entropy density, expressed as dilogarithm functions, takes a Shannon-like expression. The phase structure and the phase boundaries are obtained from the analysis of the entropy density by exploring the non-differentiable points of the entropy density. In the asymptotic regime of the isoradial class, there are universal features for the thermodynamical observables and the phase structure. The universal features in the phase structure of the isoradial class, which is expected at criticality, is rigorously studied using the hyperbolic 3-manifolds. 
Analytic properties of the special functions such as Bloch-Wigner determine the analytic/non-analytic behaviour of the entropy density of the isoradial quiver.

In summary, the hyperbolic volumes, the Mahler measures and their relations to dilogarithm functions, enable us to rigorously study the quivers in the asymptotic limit and obtain a possible phase structure of these theories. 
Finally, some concrete examples of the isoradial quivers are considered and explicit and numerical results for the entropy density, growth rate and phase structure are obtained.  

The rest of the paper is organized as following. In chapter two, we explain the crystal melting model and dimer model for the BPS sector of the toric quiver gauge theories and review the general construction of the isoradial quivers and related hyperbolic 3-manifolds. In chapter three, we develop the asymptotic analysis of the isoradial quivers. First, we set up the backgrounds and basic methods from Mahler measure theory and hyperbolic 3-manifolds. Then, using these methods and results, we obtain the free energy, BPS entropy density and growth rate. The phase structure of the quivers are studied as a result of the analysis of the entropy density. In chapter four, we consider some concrete examples of the isoradial quivers and obtain explicit results. Finally, the possible issues and directions for further studies are discussed. 

\section{BPS Counting in Isoradial Quivers}
In this section, we briefly review the construction of the BPS states, i.e. D6-D2-D0 bound states on the singularities of the Calabi-Yau threefolds, associated with the isoradial crystal melting models, and the quiver gauge theories. We will briefly introduce and review the properties of the isoradial quivers and their duals, the isoradial dimer graphs. In this work, our emphasis is on the isoradial quiver and the possible implications and consequences that the isoradiality brings into the problem of the asymptotic counting of the BPS states.

\subsection{The BPS States and Crystal Melting}
In this part, we review the construction of the crystal melting model associated with the isoradial quivers.

In the context of type-IIA string theory, we study the D6-D2-D0 bound states of D-branes in toric Calabi-Yau threefold $X$, associated with the isoradial dimer models and quivers \cite{Nish, Ya-Oo2}. The D6 brane is a noncompact brane filling the entire $X$, D2's are wrapping the two-cycles of the Calabi-Yau and D0 branes are point particles. An interesting problem is the counting of D2-D0 bound states inside the D6. On the other hand, the D-brane bound states are $\frac{1}{2}$-BPS bound states of the $\mathcal{N}=2$, supersymmetric gauge theory in four dimensions. The BPS states are charged with the $n$ number of D0 and $l$ number of D2 branes. The number of such BPS states with the charge $(n,l)$ is called BPS index and denoted by $\Omega (n,l)$. To compute the BPS index, we need to define the BPS generating function as,
\bea
\label{BPS gf}
\cZ_{\textit{BPS}}=\sum_{n, l} \Omega(n, l)\ Q^l\ q^n,
\eea
where $q$ is the fugacity factor for D0's and $Q=\{Q_I\}$ is the set of fugacity factors associated with D2 branes and $Q^l=\prod_{I} Q_I^{l_I}$ with $I$ runs over the two-cycles. The fugacity factors have geometric interpretations in topological strings, where D0 fugcity is related to coupling constant via $q= e^{-g_s}$, and D2 fugacities are related to K\"{a}hler parameters via  $Q_I=e^{t_I}$.

There are different approaches to count the BPS states and compute the BPS index. A plausible approach is using the statistical lattice model associated to dimer and quiver graphs. This lattice model is called crystal melting model and introduced in the context of string theory in \cite{Ok-Re-Va}. The construction of the crystal melting model in string theory and for a general toric quiver is developed later on in \cite{Ya-Oo2}, using the path algebra and its modules. Roughly speaking, the construction of the crystal is made by piling of the atoms of the crystal on the nodes of the quiver and the number of the atoms is related to the number of the D0 and D2 branes and equivalently to the rank of the gauge groups of the quiver. The crystal melting construction of the BPS states leads to the fact that the BPS bound states are corresponding to the crystal configurations. Moreover, there is a bijection between the dimer coverings of the toroidial dimer model and the configurations of the toric crystal melting model. These correspondences ofcourse continue to hold in the asymptotic limit, i.e. the large charge limit. Thus, the BPS index and its asymptotics can be studied using the analytic and combinatorial tools from the dimer model and crystal melting model. The first step in the counting problem and its asymptotic analysis is to define the generating function for the degeneracy of the crystal configurations $\pi_i$ from the set of all possible molten configurations denoted by $\pi$,
\bea
\label{CM gf}
\cZ_{\textit{CM}}= \sum_{\pi_i\in \pi}\prod_{i=0}^{L-1} q_i^{|\pi_i|},
\eea
where $L$ is the number of the colors, indicating different types of the atoms in the crystal, and the periodic fugacity factor $q_i$ associated with $\pi_i$, and $|\pi_i|$ is the number of the atoms in this configuration. Using the following relations between the parameters of the BPS system and crystal model; $q= q_0\ q_1\ ...\ q_{L-1}, \quad Q_I=q_i \quad (I=i=1, ..., L-1)$, one can observe that the generating functions match as $\cZ_{\textit{CM}}= \cZ_{\textit{BPS}}$ and thus the BPS index $\Omega(n,l)$ is equal to the degeneracy of the molten crystal configurations where $n$ is the total number of atoms removed from the crystal and $l_I$ is the relative numbers of different types of the atoms.

The goal of this paper is to study the asymptotics of the BPS index in the large charge limit, $n\rightarrow \infty$, in the isoradial quivers. In our previous study \cite{Zah}, the BPS entropy density and the growth rates of the toric quiver are computed using the dimer model and via the Mahler measure technology. In this work, we adopt similar techniques from the Mahler measure, the hyperbolic 3-manifolds and related powerful analytic methods around the dilogarithm function.

\subsection{Isoradial Quiver/Dimer Model and Hyperbolic Polyhedra}
%[Read and edit(change) the text]

In this part, we review the construction and properties of the isoradial dimer model and quivers and the hyperbolic manifold built on the isoradial quiver. The quiver is a planar oriented graph denoted by $Q=(Q_0, Q_1, Q_2)$, a set of nodes $Q_0$ and arrows $Q_1$ with oriented cycles $Q_2$, i.e. the faces of quiver. The head and tail of an arrow $a_i$ is denoted by $h(a_i)$ and $t(a_i)$, respectively.
Periodic quiver $Q$ is isoradial or isoradialy embedded in the plane if all the faces $Q_2$ which are polygons are inscribed in circles with radius one. In the isoradial quiver, all the arrows $a_i\in Q_1$ standing on an arc $\alpha_{a_i}\in (0,\ 2\pi)$ and for every cycle $c\in Q_2$, we have $\sum_{a_i} \alpha_{a_i}= 2\pi$.

The quiver $Q$ has a consistent $R$-charge if there is a map $R: Q_1\rightarrow (0,\ 2)$, such that for all cycles $(a_1 a_2 ... a_k)\in Q_2$ and  for all vertices $v\in Q_0$, we have
\bea
\label{iso cond}
\sum_{i} R_{a_i}=2,\quad \sum_{h(a_i)=v} (1-R_{a_i}) + \sum_{t(a_i)=v} (1-R_{a_i})=2.
\eea
The above equations are equivalent to the geometrical isoradiality condition and stating that the isoradial quiver has consistent $R$-charge 2 and vanishing $\beta$-function, \cite{Ha-Ve}.
The relation between the isoradiality and $R$-charge consistency is that a quiver $Q$ can be isoradially embedded if it admits a consistent $R$-charge. In other words, the relation between the consistent $R_{a_i}$-charges and the arcs $\alpha_{a_i}$ is obtained in \cite{Ha-Ve}, as
\bea
\pi R_{a_i}=\alpha_{a_i}.
\eea

The dual graph of the quiver is called the dimer graph. Similarly, the dimer graph is isoradially embedded if all faces which are polygons are inscribed in circles with radius one. To each edge of the dimer graph $e_i\in E$, one assign an angle $\theta_{e_i}\in (0,\ 2\pi)$. By connecting nodes of the dimer graph to the centers of the dimer graph, i.e. nodes of the quiver, we have the tiling of the plane by the rhombi with side of length 1. This is called rhombus tiling. Dimer edges and quiver arrows are perpendicular diagonals of the rhombus tiling. The two notions of isoradiality in dimer and quiver graphs are related by the complementary arcs for the dual graphs,
\bea
\alpha_{a_i}= \pi- \theta_{e_i}\quad \textit{if}\quad e_i=\breve{a}_i,
\eea
where the edge $e_i$ is the dual arrow $a_i$ of the quiver. For simplicity we will denote $R_{a_i}$ by $R_i$, and $\alpha_{a_i}$ by $\alpha_i$ in the following sections and will normalize the angles and the $R$-charges by a factor 1/2 such that $\sum_i R_i=1$.

Further analysis of the consistency of the quiver gauge theories, related to dimer model is discussed in \cite{gul}. For a comprehensive review of the geometric aspects of the dimer model and related quivers, consult with \cite{boc}.
In the next part, we will explain the statistical mechanics of the dimer model and its relation to quiver gauge theories.

\subsubsection*{Characteristic Polynomial of Isoradial Quivers}
Given a toric Calabi-Yau manifold $X$ there is an associated Newton polynomial (NP), \cite{Bo}, 
\bea
\label{CY NP}
P(z,w)= \sum_{(n,m)} a_{(n,m)}\ z^n\ w^m,
\eea
where $z=e^{2\pi \mathrm{i}\theta_1},w=e^{2\pi \mathrm{i}\theta_2}$ are the two coordinates of the torus, $(n,m)$ are the vertices of the toric diagram and $a_{(n,m)}$ are associated with the K\"{a}hler moduli $t_{(n,m)}$ of $X$. The Newton polygon or the toric diagram of $P$ is the convex hull in $\R^2$ of the set $\{(n,m)\ |\ a_{(n,m)}\neq 0\}$. The NP encodes all the geometric data in $X$ and the associated quiver. However, the thermodynamic information of the quiver can be encoded in an extension of the NP called the characteristic polynomial of the isoradial dimer. This is defined as the sum over the dimer covers $\cC$ in the set of dimer coverings $\cM$ on the dual graph of the quiver, \cite{Ken},
\bea
\label{dimer NP}
P(z,w; \{\varepsilon_i \}) = \sum_{\cC \in \cM} e^{-\varepsilon_\cC} z^{h_x} w^{h_y} (-1)^{h_x h_y},
\eea
where $h_x, h_y$ are the height changes in the dual graph and given by the vertices of the toric diagram $(h_x, h_y)= (n,m)$, and the energy of the quiver is the sum of the energies of the dimer edges,  $\varepsilon_\cC= \sum_{i\in \cC} \varepsilon_i$.  The energy of the dimer edge is related to the conformal dimension and the $R$-charge of the associated field of the quiver. In the isoradial dimer, the energy of the dimer edge is the logarithm of the length of the edge of the quiver. 
As we will observe, the characteristic polynomial determines the statics and dynamics of the asymptotics of the quiver. The set of data encoded in the NP, completely determine the thermodynamics of the (isoradial) quivers. The exponents $(n,m)$ of the NP are fixed parameters set by the underlying Calabi-Yau geometry provide some fixed property of the quiver's thermodynamics such as phase boundaries. The coefficients $a_{(n,m)}$ of the NP are the dynamical parameters and control the dynamical properties of the quiver and thermodynamical observables such as entropy density.

From NP in Eq. \eqref{CY NP}, we can define a NP, analogous to Eq. \eqref{dimer NP}, associated with the generalized quiver polygon with edge lengths $a_i$, as
\bea
P(z,w; \{a_i\})= \sum_{i} a_{i}\ z^{n_i}\ w^{m_i},
\eea
where $i$ runs over the boundary vertices of the toric diagram with component $(n_i, m_i)$ and $\{a_i\}\equiv (a_1, a_2, ..., a_n)$. In fact, to each term in NP in Eq. \eqref{CY NP}, we can associate a coefficients $a_i$ which is the edge lengths of the quiver. This way, the quiver is a polygon with the same number of sides as toric diagram.
In the isoradial quiver, in consistency with Eq. \eqref{dimer NP}, we have the following relations $a_i=\nu_i=e^{-\varepsilon_i}= 2\sin\pi R_i$, where $\ni_i$ are the statistical weights on the edges of the dimer model. Indeed, the edge lengths of the quiver is given by the $R$-charges of the fields associated with the quiver edges.  In the statistical picture, the quiver edge lengths are the statistical weights and their logarithm are energies of the dimers i.e. the fugacity factors associated with $R$-charges. For more convenience, we will denote the fugacity factors by $x_i$, defined from $\log a_i= x_i$. For simplicity, we will denote $P(z,w; \{a_i\})$ by $P(z,w)$, unless it is necessary to keep the dependence on $a_i$ explicit.

\subsubsection*{Hyperbolic Polyhedra, Ideal Triangulation and Isoradiality}

In this part we explain the construction of the hyperbolic 3-manifolds built on the isoradial quiver polygon and its triangulations into tetrahedra. we briefly discuss the conditions for the triangulation and also its relation to the isoradilaty conditions. For more details, see the Appendix and references therein.

Let $N$ be an ideal polyhedron, a hyperbolic compact, 1-cusped and finite volume 3-manifold, with torus boundary, constructed over the quiver polygon with edge lengths $a_i$.
In other words, the projection of the ideal polyhedron on the plane is a cyclic polygon in $\partial\H$, i.e. the isoradial quiver, explained in the previous section.
The ideal triangulation of $N$, into finite union of ideal tetrahedra $T(z_j)\equiv T_j$ characterized by  shape parameters $z_j$, and with vertices at $0,1,\infty, z_j$, and the volume of the polyhedron are as follow, see the Appendix for more details,
\bea
M = \bigcup_{j=1}^n T_j, \quad  \textit{Vol}(N)= \sum_j D(z_j).
\eea
The necessary and sufficient conditions for ideal tetrahera $T_j$ and their shape parameters $z_j$ to glue around each edge of the triangulation and form the ideal polyhedron $N$ is known as gluing equations,
\bea
\prod_{i=1}^n z_i^{a_{i,j}} (1-z_i)^{a'_{i,j}}= \pm 1, \quad \textit{for}\ j= 1, ..., n,
\eea
where $a_{i,j},\ a'_{i,j}$ are integers depending on $N$. These algebraic equations has a solution $(z_1, z_2, ..., z_n)$ with $\Im z_j > 0$, for all $j$ and it is called geometric solution. Moreover, there are other conditions dealing with fitting properties at the cusps of the $N$, called completeness equation,
\bea
\prod_{i=1}^n z_i^{c_i} (1-z_i)^{c'_i}= \pm 1, \quad \prod_{i=1}^n z_i^{d_i} (1-z_i)^{d'_i}= \pm 1,
\eea
where $c_i, c'_i, d_i, d'_i$ are integers depending on $N$. In fact, the metric of the hyperbolic 3-manifolds is complete iff $(z_1, z_2, ..., z_n)$ solves the completeness equations. 

In the triangulation of the hyperbolic polyhedra, the necessary and sufficient conditions for the existence of the solutions of the gluing equations are discussed in \cite{Bo-Sp}, which in our notation are
\bea
\sum_{f \in Q_2} 2\pi= \sum_{a_i\in Q_1} 2(\pi- \alpha_{a_i}), \quad \sum_{f \in Q'_2 }2\pi < \sum_{a_i\in Q'_1} 2(\pi- \alpha_{a_i}),
\eea
where $Q'_2$ is a nonempty subset of faces and $Q_2\ne Q'_2$, and $Q'_1$ is the set of all edges incident with any face of $Q'_2$. In fact, the above gluing conditions follow from the isoradialty conditions in Eq. \eqref{iso cond} by taking the sum over $f\in Q_2$ and $f\in Q'_2$. In essence, these equations imply that around every edge, the gluing should be such that the sum of the angles sum up to $2\pi$. Thus, the triangulation and gluing equations are equivalent to the isoradiality conditions for the isoradial quiver. 

\subsubsection*{NP of Isoradial Quiver and A-Polynomial}
The NP of isoradial quiver are genus zero curves. Furthermore, as the isoradial quivers can be triangulated and they satisfy the gluing and completeness equations, their NP are similar to to A-polynomials and their closed cousines H-polynomials, for an introduction to A-polynomials and their Mahler measure see \cite{sen}. 
A-polynomials are special type of NP which satisfy the algebraic analogue of ideal triangulation,
\bea
z\wedge w = \sum_{j=1}^n r_j\ z_j\wedge (1-z_j),
\eea
for some $r_j\in \mathbb{Q}$. First, we make an observation about the relation between the A-polynomials and the Newton polynomials of the isoradial quivers.
The NP of isoradial, like the A-polynomials, satisfies the triangulation equations and can be triangulated. They also have the property that their Mahler measure are the volumes of some hyperbolic 3-manifolds. 
Thus the NP of the isoradial quivers shares the same properties with the A-polynomials.

In fact, the A-polynomial defines the NP associated with the isoradial quivers and vice versa.
To each A-polynomial we can associate a cyclic polygon where the number of the edges is equal to the number of the terms in the A-polynomial and the length of each edge $a_i$ is the coefficient $a_{(n,m)}$ of the associated term in the A-polynomial. If all of the $a_i$'s are the same we can scale them to one and obtain a normalized A-polynomial with all the coefficients equal to one. Particularly, this one is of special interest since, as we will see, it maximizes the volume of the ideal hyperbolic polyhedra built on the cyclic polygon and consequently maximizes the Mahler measure of the A-polynomial. 
 
\section{Asymptotic Analysis of Isoradial Quivers}
As we obtained in our previous study \cite{Zah}, the growth rate (and the BPS entropy density) are obtained from the surface tension of the dimer model and explicitly computed using the (generalized) Mahler measure of the NP of the quiver. The free energy of the quiver theory is obtained from the limit shape of the crystal model in terms of the Ronkin function.
In the special case of isoradial quivers, the Newton polynomial is of special type, namely the A-polynomials. The Mahler measure of A-polynomials has been studied specially for their relations to volumes of hyperbolic 3-manifolds. We use the language and techniques from the hyperbolic geometry to study the asymptotics of isoradial quivers. To do this, first we compute the Mahler measure of the isoradial quivers and then apply this to compute the thermodynamical quantities of the isoradial quiver such as free energy, entropy density and growth rate, using the physical parameters of the quiver, i.e. the R-chatrges and associated fugacities. This leads to the exploration of the possible phase structure in the isoradial quivers and its universality features.

In the next part, first we review the results for the asymptotic counting of BPS states using the dimer model and crystal melting and associated Mahler measure.
\subsection*{Crystal Quiver Asymptotics and Mahler Measure}
\label{quiver crystal asym}

In this part, first we review the results obtained in \cite{Zah} about Mahler measure method in study of the asymptotic of the toric quivers. In our previous work, we established the connection between the asymptotics of the BPS sector and the generalized Mahler measure theory. Building on our previous results, we introduce related techniques from hyperbolic 3-manifolds to study the asymptotics of the isoradial quivers to obtain explicit results for the thermodynamical observables.

For the dimer model, defined on the $M$ by $M$ cover of the torus, using the asymptotic techniques, the BPS partition function in the large $M$ limit is obtained in \cite{Ya-Oo}, as
\bea
\label{BPS gf 2}
\cZ_{\textit{BPS}}\sim\exp{\Bigg(M^2\left(\int_{0}^1 -\sigma(s, t)\ dx\ dy- g_s M \int_{0}^1 N(x,y)\ dx\ dy\right) \Bigg)},
\eea
where $N(x,y)$ is the profile function of gauge groups of the quiver, obtained in the asympototic limit by the appropriate scaling and is given by the Ronkin function,
\bea
\label{Ronkin}
N(x,y)=\frac{1}{(2\pi \mathrm{i})^2} \int_{|z|=e^x, |w|=e^y} \log|P( z, w)|\ \frac{dz}{z}\frac{dw}{w},
\eea
and the surface tension $\sigma(s,t)$ is given by the Legendre dual of the Ronkin function.

For the computational purposes, one can formulate the asymptotics in terms of the Mahler measure. The Mahler measure of the Newton polynomial $P(z,w)$, is defined by
\bea
m(P)= \frac{1}{(2\pi \mathrm{i})^2} \int_{|z|=1, |w|=1} \log|P(z, w)|\ \frac{dz}{z}\frac{dw}{w},
\eea
and similarly the generalized Mahler measure on the torus is defined by
\bea
m_{(a,b)}(P)= \frac{1}{(2\pi \mathrm{i})^2} \int_{|z|=a, |w|=b} \log|P(z, w)|\ \frac{dz}{z}\frac{dw}{w}=\frac{1}{(2\pi \mathrm{i})^2} \int_{|z|=1, |w|=1} \log|P(az, bw)|\ \frac{dz}{z}\frac{dw}{w}.
\eea
The generalized Mahler measure  reduces to the original one at $a=b=1$. At $a=e^x$ and $b=e^y$, the generalized Mahler measure is the Ronkin function, as we have
\bea
\label{Ron mah}
N(x,y)=m\Big(P(e^x z, e^y w)\Big)=m_{e^x,e^y}(P).
\eea
The surface tension $\sigma(s,t)$ is the entropy density of the quiver and is given by the Legendre transform of the profile,
\bea
\label{leg trans}
\sigma(s,t)=\mathcal{L}[N(x,y)]= -N(x,y)+ x\ s+ y\ t.
\eea
In terms of the Mahler measure we have
\bea
\label{Leg dual}
\sigma(s,t)&=& \mathcal{L}(m_{e^x,e^y}(P))\nonumber\\
&=& -m_{e^x,e^y}(P) + xs + yt,\nonumber\\
(s,t)&=&\nabla m_{e^x,e^y}(P),
\eea
where the last line is implied by the Legendre duality.

Using the Legendre transform properties and putting the total derivative term to zero, the free energy of the quiver is obtained in \cite{Ya-Oo}, as the integral of the profile function over the Amoeba $\cA=\{\ (\log|z|,\log|w|)\ |\ P(z,w)=0\ \}$ (for a review on the Amoeba and related topics, see \cite{Mik}),
\bea
\label{dimer free energy}
\mathcal{F}_{\textit{BPS}}=\log \cZ_{\textit{BPS}}= -\frac{1}{g_s^2} \int_\cA N(x,y)\ dx\ dy = -\frac{1}{g_s^2}\int_\cA  \mathcal{L}[\sigma(s,t)] \ dx \ dy=-\frac{1}{g_s^2} \int_{\mathcal{A}} m_{e^x,e^y}(P)\ dx \ dy.
\eea

The BPS growth rate of the quiver is obtained from the maximum value of the entropy density, and it is given by the generalized Mahler measure at $a=b=1$, \cite{Zah},
\bea
\lim_{n\rightarrow \infty}\log \Omega(n) = -\sigma(s^*, t^*)^{1/3} n^{2/3}= N(0,0)^{1/3}n^{2/3}= m_{(1,1)}(P)^{1/3} n^{2/3}.
\eea
In \cite{Zah}, the above general formalism is applied to some fundamental examples of the toric quivers and explicit results are obtained. In this paper, our goal is to extend the asmyptotic analysis of the BPS sector to the class of isoradial toric quivers and obtain the general explicit results in this class. In fact, the main goal of this work is to apply the above result to the case of isoradial quivers and obtain explicit result in this case. This will be done by introducing a new formulation of the problem in terms of hyperbolic 3-manifolds.

\subsubsection*{Dimer Model vs Hyperbolic Parameters}
One of our goals in this article is to introduce and develop a physical parameterization for the thermodynamics of the quivers in terms of the $R$-charges of the quiver fields. We generalize the dimer model two-variable formalism in the BPS sector of the toric quiver to the multivariate hyperbolic parameterization suitable in the class of isoradial quivers. 
As we will explain throughout the paper, working in this setup has some advantages. Especially, since the isoradial dimer weights are proportional to the lengths of the edges of the quiver, thus we can use this property to connect the asymptotic of the isoradial quivers to the Mahler measure and hyperbolic 3-manifolds, in a natural way.

Using the hyperbolic parametrization, we can setup a physical formalism to study the thermodynamics of the isoradial quivers and obtain the entropy density, growth rate, free energy and the phase structure in terms of the parameters of the hyperbolic 3-manifolds and the $R$-charges of the quiver. In principle, for any quiver, it is always possible to transform the results into the dimer parametrization and obtain an intuitive picture in terms of crystal model. This might be considered as an advantage but mathematically equivalent to the hyperbolic parameterization. A more concrete explanation of the relations between the two parameterization is explained in section \ref{reduction}.

Before we proceed to the hyperbolic formulation of the asymptotics of the isoradial quivers, we need to review some mathematical backgrounds and results about the Mahler measure and hyperbolic 3-manifolds. 

\subsection{Isoradial Quivers, Mahler Measures and Hyperbolic 3-manifolds}
As we obtained in \cite{Zah}, the BPS entropy density and the growth rates are obtained from the Mahler measure and its generalization on the torus. In the case of isoradial quivers, the NP of the quivers shares many properties with the A-polynomials. Thus, in order to study the asymptotics of the isoradial quivers, we need to study the Mahler measures of the A-polynomials. 
In this part, first we review the relation of the Mahler measure of A-polynomials and volumes of the hyperbolic 3-manifolds, in the simplest example. Then, based on this relation, and using the generalization of the Ronkin function and its Legendre property, we obtain the Mahler measure of the isoradial quivers in terms of the generalized Legendre transform of the hyperbolic volumes.

\subsubsection*{Mahler Measure and Hyperbolic Volumes}
\label{mah hyp}
Let us first consider the normalized NP of isoradial quiver with all the coefficients set to one $\{a_i=1\}$. Extending the Jenkins formula to the Mahler measure of the NP with two variables, Boyd and Rodriguez-Villegas computed the Mahler measure of the A-polynomials \cite{bo-Ro}.
Using the solutions $w_i$ of $P(z,w)=0$, the NP can be factorized as,
\bea
P(z,w) = w_0(z) \prod_{i=1}^d (w-w_i(z)).
\eea
Then the Mahler measure of the NP can be decomposed as \bea
m(P)&=& m(w_0)
+ \frac{1}{(2\pi \mathrm{i})^2} \int_{|z|=1, |w|=1} \sum_{i=1}^d\log|w-w_i|\ \frac{dz}{z}\frac{dw}{w}\nonumber\\
&=& m(w_0)+\frac{1}{2\pi \mathrm{i}} \sum_{i=1}^d\int_{|z|=1, |w_i|\geq 1} \log|w_i|\ \frac{dz}{z}\nonumber\\
&=&  m(w_0)-\frac{1}{2\pi} \sum_{i=1}^d\int_{|z|=1, |w_i|\geq 1} \eta(z,w_i),
\eea
where $\eta(z,w_i)=\log|z|\ d \arg w_i- \log|w_i|\ d \arg z= - \log|w_i|\ d \arg z$ for $z=e^{2\pi \mathrm{i} \arg z}$ on the torus, and we used $d\arg x = \Im (\frac{dx}{x})$. For cyclotomic polynomials which all the roots of $w_0(z)$ lie on the unit circle, we have $m(w_0)=0$. 

For the normalized NP of isoradial quivers which are genus zero curves and for A-polynomials of the 1-cusped hyperbolic 3-manifolds, we can solve the gluing equations and thus using the relation between the volume form and Bloch-Wigner function (see the Appendix), we have
\bea
z\wedge w_i&=& \sum_{i_k} r^*_{i_k} z^*_{i_k}(1-z^*_{i_k}),\nonumber\\
\eta(z,w_i)&=& \sum_{i_k}\eta(z^*_{i_k}, 1-z^*_{i_k})=\sum_{i_k} d D(z^*_{i_k})= d(\sum_{i_k}  D(z^*_{i_k}))= d\textit{Vol},
\eea
to write
\bea
m(P)&=& \frac{1}{2\pi} \int_\gamma d\textit{Vol}= \frac{1}{2\pi} \int_{\partial \gamma} \textit{Vol}= \sum_{l} \textit{Vol} ((z_l, w_l)^1)- \textit{Vol} ((z_l, w_l)^2)\nonumber\\
&=&-\frac{1}{2\pi} \sum_i \sum_{i_k} r^*_{i_k} D(z^*_{i_k})\rvert_{\partial \gamma},
\eea
where $\gamma$ is an oriented path in $\{(z,w)|\ |z|=1 , |w_i|\geq 1\}$, $(z_l, w_l)^1, (z_l, w_l)^2$ are the boundary points of the $\gamma$, and we used the Stokes Theorem in the first line with the sum over the volumes of hyperbolic 3-manifold under different hyperbolic metrics. Thus, we observe that the Mahler measure 
are related to the volume of the hyperbolic 3-manifold. The relation between the Mahler measure of A-polynomials and the volume of the hyperbolic 3-manifolds is proposed in \cite{bo-Ro,boy} and developed in \cite{Lal-hyper}, based on initial numerical observation in \cite{boy-L} and the above computations provide possible explanation of this relation. We explain further this relation in the case of isoradial quivers. As we will describe in the next part, the Mahler measure of isoradial quivers with arbitrary edge lengths $\{a_i\}$ is the volume of the hyperbolic ideal polyhedron $N$ built on the quiver cyclic polygon plus logarithmic terms,
\bea
\pi m(\{a_i\})= \textit{Vol}(N) + \textit{log-terms}.
\eea
The Mahler measure and the volume of the ideal polyhedron is maximized in the isoradial quivers with the same edge lengths $\{a_i=a^*_i=1\}$, which corresponds to the regular cyclic polygon. Lets denote the corresponding ideal regular polyhedron by $N^*$, then, in consistency with the above result, we have, 
\bea
\pi m(\{a^*_i\})= \textit{Vol}(N^*).
\eea

\subsubsection*{Mahler Measure of Isoradial Quiver}
In the previous part, we have computed the Mahler measure of the normalized NP. In this section we turn on the coefficients $a_i$ and compute the Mahler measure.
As we observed in \cite{Zah}, the generalized Mahler measure, on the arbitrary torus, of the NP of the quiver is the Ronkin function that depends on the torus parameters $a,b$. It is natural to generalize the Ronkin function to depends on $\{a_i\}$. Using the Legendre duality relation between the generalized Mahler measure and the entropy density in Eqs. \eqref{Ron mah} and \eqref{leg trans},
\bea
\label{Leg mah}
m_{(a,b)}(P)= -\sigma (\nabla_{\textit{log}}  m_{(a,b)})+ \frac{\partial m_{(a,b)}}{\partial \log a} \log a + \frac{\partial m_{(a,b)}}{\partial \log b} \log b,
\eea
where $\nabla_{\textit{log}}=(\frac{\partial}{\partial \log a}, \frac{\partial}{\partial \log b})$, and the fact that the maximized entropy density is given by the generalized Mahler measure at $a=b=1$, as $\sigma^*= m_{(1,1)}(P)$, we can generalize the Legendre duality relation in  Eq. \eqref{Leg mah} and write the Mahler measure (the generalized Ronkin function) of the NP of Isoradial quiver as a cyclic polygon with edge lengths $a_j$ and dihedral angles $\alpha_j$,  as
\bea
\label{gen ronk}
\pi m(\{a_j\})=\textit{Vol}(N)+ \sum_j \alpha_j \log a_j= \sum_{i\in T} D(z_i)+ \sum_{j} \alpha_j \log a_j,
\eea
where the first sum is over the triangles in the triangulation $T$ and the second sum is over the sides of the poygons indexed by $j$.
The above mathematical result will be used in this article in the explicit computations of the Mahler measure of the NP of the isoradial quivers, and can be seen as the generalization of the results in \cite{bo-Ro}.

\subsubsection*{Relation to Isoradial Dimer Model}
The above result for the generalized Ronkin function in Eq. \eqref{gen ronk} can be obtained from another perspective, using the statistical mechanics of the toric isoradial dimer model.
First, it is obtained in \cite{Ke-Ok-Sh}, that  the partition function of the dimer model on the toric bipartite planar graph is given by the Mahler measure,
\bea
\label{dim mah}
\log \cZ^{\textit{toric}}=\frac{1}{(2\pi \mathrm{i})^2} \int_{|z|=1, |w|=1} \log|P( z, w)|\ \frac{dz}{z}\frac{dw}{w}.
\eea
In presence of the magnetic field $(B_x, B_y)$, the above partition function becomes the Ronkin function $N(B_x, B_y)$.

On the other hand, the partition function of the isoradial dimer model on an isoradial periodic bipartite planar graph is obtained in \cite{Ken-iso,de}, as
\bea
\log \cZ^{\textit{iso}}= \sum_{e\in E} \frac{1}{\pi} L(\theta_e)+ \frac{\theta_e}{\pi} \log (2\sin \theta_e),
\eea
where $\theta_e$ is the rhombus half-angle associated to the edge $e$ in the set of all edges $E$.
Combining the above results for the partition function of the toric dimer model and the results for the partition function of the isoradial dimer model, we obtain the partition function of the toric isoradial dimer model. In fact, the free energy of the dimer model on an isoradial periodic bipartite planar graph $G$ is obtained from the partition function of the toric dimer model on the univeral cover of a finite isoradial graph on a torus, $G_M=G/ M\mathbb{Z}^2$,  for an integer $M>0$, as
\bea
\log \cZ(G)\coloneqq\lim_{M\rightarrow \infty} \frac{1}{M^2} \log Z(G_M) = \sum_{e\in E} \frac{1}{\pi} L(\theta_e)+ \frac{\theta_e}{\pi} \log (2\sin \theta_e).
\eea
Using the duality of the dimer model and quivers, the above result can be seen as the Mahler measure of the isoradial quivers, and indeed consistent with Eq. \eqref{gen ronk}. 

Moreover, the surface tension of the isoradial dimer model is obtained from the Legendre transform of the free energy and is given by the volume of the hyperbolic polyhedron,
\bea
\sigma(\{\theta_e\})=\mathcal{L}[\log \cZ(G)]=-\frac{1}{\pi}\sum_e L(\theta_e).
\eea
For further statistical mechanical results on isoradial dimer model, see \cite{bout}.

Now, if we use the triangulation of the polygon, we can observe the equality between two expressions for the entropy density in terms of the Bloch-Wigner and Lobachevsky function,
\bea
\sum_{i\in T} D(z_i)= \sum_{i\in T} \big(L(\theta_i^1)+L(\theta_i^2)+L(\theta_i^3)\big),
\eea
where $\theta_i^1=\arg z_i$, $\theta_i^2=\arg (1-\bar{z}_i)$ and $\theta_i^3=\arg (\frac{1}{1-z_i})$.
%and the cancellation of the contributions from the opposite angles in front of the internal edges. 
In a triangulation of isoradial quiver polygons we have, $\theta_i^3=\pi-\theta_j^3$, for any $(i,j)$ pair of opposite triangles that share an internal edge. Thus, using this and Eq. \eqref{gen ronk}, we can write the Mahler measure in terms of the Lobachevsky function,
\bea
\label{Mah iso 1}
\pi m(\{a_i\}) = \sum_{i} L(\alpha_i)+ \alpha_i \log a_i= \textit{Vol}(N)+\sum_{i} \alpha_i \log a_i.
\eea
To relate the generalized Ronkin function to the entropy density, we can use the multivariate Legendre transformation,
\bea
\label{Mah iso 2}
m(\{a_i\}) =-\sigma(\{\alpha_i\})+ \sum_{i} \frac{\alpha_i}{\pi} \log a_i.
\eea
The observations in this section about the connections between Mahler measure and dimer model, and Mahler measure and hyperbolic volumes may shed light on the  mysterious connections between isoradial dimer model and hyperbolic volumes observed in \cite{Ken-iso}.  

\subsection{Free Energy of the Isoradial Quivers: $R$-Charge Parameterization}
As we explained in the beginning of the section \ref{quiver crystal asym}, the BPS entropy density and the free energy of the quiver is obtained in terms of the Mahler measure. Specialization to the isoradial quivers leads us to a specific Mahler measure, namely the Mahler measure of the A-polynomials. In the class of isoradial quivers, due to the connection between the Mahler measure of the A-polynomials and the hyperbolic ideal polyhedra, we can obtain explicit results. In fact, hyperbolic geometry provides the explicit results about the asymptotic and the phase structure of the model. 

Having explained the change of parameterizations and the relations between the hyperbolic parameters and the isoradial quivers, we continue towards the formulation of the asymptotic analysis of the isoradial quivers in terms of the hyperbolic parameters. In this part, we derive the asmyptotic analysis of the isoradial quivers in terms of the physical quantities, the $R$-charge $R_i= \frac{\alpha_i}{\pi}$, and associated fugacities $x_i=\log a_i$, $x_i=\log \sin \pi R_i$. In this part, we introduce a multivariate generalization of the asymptotic analysis of the toric quivers explained in \cite{Zah} and reviewed in the beginning of the section (\ref{quiver crystal asym}). This is the hyperbolic/$R$-charge parameterization of the asymptotics of the isoradial quiver gauge theories. The relation between the BPS index of toric isoradial quiver gauge theories and hyperbolic 3-manifolds has been proposed and studied in \cite{Te-Ya,Ya-YBE}. As a natural generalization of the two-variable asymptotic analysis, we start with the asymptotics of the partition function parameterised in the hyperbolic parameters,
\bea
\label{BPS gf 2}
%\lim_{n\rightarrow \infty}
\cZ_{\textit{BPS}}\sim\exp{\left(M^d\int\prod_{i=1}^d \cJ d x_i  (-\sigma(\{R_i\}) - g_s M m(\{x_i\}))\right)},
\eea
where $x_i= \log a_i$, and $\cJ$ is the Jacobian of the transformation from two-variable to $d$-variable, with the scaling dimension $[\cJ]= M^{2-d}$. Using the Legendre duality between the entropy density and the generalized Ronkin function, we have
\bea
m(\{x_i\})=-\sigma(\{R_i\})+ \sum_i R_i x_i,
\eea
and Legendre transformation implies,
\bea
\frac{\partial \sigma}{\partial R_i}= x_i, \quad \frac{\partial m}{\partial x_i}= R_i.
\eea
The generalized Ronkin function in terms of the $R$-charges, can be obtained from Eq. \eqref{Mah iso 1},
\bea
m(\{R_i\})=\frac{1}{\pi}\textit{Vol}(M)+ \sum_i R_i \log \sin \pi R_i= \frac{1}{\pi}\sum_{i\in T} D(z_i)+ \sum_{i} R_i \log  \sin \pi R_i,
\eea
\bea
m(\{R_i\}) = \sum_{i} \frac{1}{\pi}L(\pi R_i)+ R_i \log (2\sin \pi R_i).
\eea
By generalizing the arguments for Eq. \eqref{dimer free energy}, the partition function is obtained as integral of the Mahler measure of the quiver over the generalized Amoeba $\cA_G= \{\ (x_1, x_2, ..., x_d)\ |\ P(z,w;\{a_i\})=0\ \}$, as
\bea
\label{BPS gf 2}
%\lim_{n\rightarrow \infty}
\cZ_{\textit{BPS}}\sim\exp{\left(-1/g_s^2\int_{\cA_g}\prod_{i=1}^d \cJ d x_i\  m(\{x_i\})\right)}.
\eea
Then, we can compute the free energy of the isoradial quivers which is the genus-zero Gromov-Witten invariants of the isoradial toric quivers. In fact, as the NP of the isoradial quiver is a genus zero curve, the generalized Ronkin function
is contributing precisely to the genus-zero Gromov-Witten invariant. In other words, the free energy of the isoradial quiver is the genus-zero partition function of topological string. The isoradiality of the quiver, implies the exact and explicit expressions for the free energy, in terms of the Mahler measure, hyperbolic volume and thus the dilogarithm functions. 
After the change of variable in the dimer model representation of the free energy in Eq. \eqref{dimer free energy},
we can find the hyperbolic parameterization of the free energy,
\bea
\mathcal{F}_{\textit{BPS}}= -\frac{1}{g_s^2}\int_{\cA_G}\prod_i \cJ d x_i\  m(\{x_i\})=-\frac{1}{g_s^2}\int_{\cA_G} \prod_i \cJ d x_i\ \mathcal{L}[\sigma(\{R_i\})]=-\frac{1}{g_s^2}\int_{\cA_G} \prod_i \cJ d x_i\ \mathcal{L}[\frac{1}{\pi}\textit{Vol}(M(\{a_i\}))].
\eea
Similar result are obtained from other plausible approaches, in \cite{Ya-YBE}.

The mean curvature of the hyperbolic 3-manifold, defined by the sum over the edges of $(\pi-\alpha_i)l$ with $l$ is the hyperbolic edge length, is the internal free energy, i.e. the mean energy of the isoradial quiver, \cite{Ken-iso}, is given by
\bea
U=\frac{1}{\pi}\sum_i \alpha_i \log (2\sin\alpha_i)=\sum_i R_i \log (2\sin\pi R_i).
\eea

\subsection{BPS Entropy Density}
In this work, we give another parameterization of the entropy density namely the hyperbolic/$R$-charge parameterization. We will also compare this parameterization with the dimer model parameterization in terms of the height functions and slopes. 
We elaborate on the BPS entropy density as the Legendre dual of the Mahler measure of the NP of the isoradial quiver.
We obtain BPS entropy density and growth rate in terms of the Hyperbolic Volumes and Bloch-Wigner Function, as a function of the $R$-charges.
As we will see, the liquid phase contribution to the BPS index is the hyperbolic volume of the ideal polyhedron built of the quiver polygon and is expressed in terms of Bloch-Wigner function. In the isoradial and isoradially embedded quivers, this is the only contribution to the BPS index, since the gas phase either does not exist or is shrunk to zero.

\subsubsection*{Entropy Density, Mahler Measure and Hyperbolic Volumes}
Similar to the two-variable case in which the entropy is obtained as the Legendre dual of the Ronkin function in Eq. \eqref{leg trans}, the entropy density in hyperbolic parameterization is given by the Legendre transform of the Mahler measure of the quiver polygon,
\bea
\label{Ent R}
\sigma(\{R_i\})=\cL(m(\{x_i\}))= -m(\{x_i\}) + \sum_i R_i x_i.
\eea

%\subsubsection*{Entropy Density and Hyperbolic Volumes}
As we observed in the previous sections, the Mahler measure of the NP of quiver polygon is related to the volume of the ideal hyperbolic polyhedron built on the quiver polygon, Eq. \eqref{Mah iso 1}. Thus, the entropy density becomes the volume of the polyhedron,
\bea
\sigma(\{R_i\})= \cL(m(\{x_i\})= -\frac{1}{\pi}\textit{Vol} (N).
\eea
This is consistent with the results in isoradial dimer model, \cite{Ken-iso}. We observe that the BPS counting of the isordial quivers as well as counting in isoradial dimer model in the thermodynamic limit can be studied using the hyperbolic 3-manifolds.
Now, we can use the computation of the hyperbolic volumes in terms of the dilogarithm functions. The BPS entropy density of the isoradial quivers can be written in two ways in terms of the Bloch-Wigner function and Lobachevsky function. In terms of the Bloch-Wigner functions, as the sum over the shape parameters in the triangulation $T$,
\bea
\label{ent BW}
\sigma(\{R_i\})=-\frac{1}{\pi}\textit{Vol} (N) =-\frac{1}{\pi}\sum_{i\in T} D(z_i),
\eea
and in terms of the Lobachevsky function,
\bea
\label{ent lobach}
\sigma(\{R_i\})=-\frac{1}{\pi} \textit{Vol} (N)=-\frac{1}{\pi}\sum_i L(\pi R_i).
\eea
The next computational step would be to compute the
volume in terms of the geometrical parameters of the polyhedron and the physical parameters, $R$-charges. We will perform these computations in concrete examples in chapter 4.

\subsubsection*{Schl\"{a}fli Formula and Entropy Density}
In this part, we further explore the connections between the thermodynamics of the quiver in terms of the Legendre transform of the Mahler measure and the Schl\"{a}fli volume form in hyperbolic 3-manifolds, \cite{kel}. The idea is to present another geometric interpretation and realization for the thermodynamics of the isoradial quiver. 

As we explained in section \ref{mah hyp}, using the Jenkins formula, the Mahler measure of the two-variable NP is related to the volume differential form,
\bea
\eta(z,w)= \log |z|\ d \arg w- \log |w|\ d \arg z.
\eea
Using the triangulation decomposition $z\wedge w= \sum_i z_i(1-z_i)$ for the NP of isoradial quivers and A-polynomials, we have
\bea
\eta(z,w)=\sum_i\eta(z_i, 1-z_i)=\sum_i d D(z_i) = d (\sum_i D(z_i)) = d\textit{Vol}.
\eea

In the hyperbolic geometry, the volume differential form of the hyperbolic polyhedron is given by the following formula known as Schl\"{a}fli formula \cite{kel}, expressed in terms of logarithmic edge lengths $l_i$ and the dihedral angles $\alpha_i$,
\bea
d\textit{Vol}= -\frac{1}{2}\sum_i l_i\ d\alpha_i.
\eea
On the other hand, in the thermodynamic formalism, the entropy density, as the Legendre dual of the Mahler measure given by Eq. \eqref{Ent R}, can be written in terms of the dihedral angles and quiver edge lengths as,
\bea
\label{ent leg dual}
\sigma (\{\alpha_i\})= -m(\{a_i\})+ \sum_i \frac{\alpha_i}{\pi} \log a_i.
\eea
From the above Legendre transformation, the entropy density differential, modulo $\pi$, becomes 
\bea
\label{ent der}
d\sigma = \sum_i  \frac{\partial \sigma}{\partial \alpha_i}   d\alpha_i, \quad \frac{\partial \sigma}{\partial \alpha_i}=  \log a_i= x_i.
\eea
Assuming the relation between the entropy density and the hyperbolic volume, and comparing the Schl\"{a}fli differential volume formula with the entropy differential, we obtain
\bea
l_i=x_i= \log 2\sin \alpha_i. 
\eea
This can be considered as a thermodynamical interpretation of the Schl\"{a}li formula, from the isoradial quiver point of view. Thus, we observe that the thermodynamics of the isoradial quivers is closely related to the convex variational principle in ideal polyhedra, \cite{ho-ri,riv,fil} and this is suggestive of a possible fruitful interdisciplinary connections.

\subsection{BPS Growth Rate}
\label{BPS Growth Rate}
In this part, we compute the BPS growth rate of the isoradial quiver. As we observed in \cite{Zah} and reviewed in section \ref{quiver crystal asym}, the BPS growth rate is computed via the extremization of the BPS entropy density.
\bea
\lim_{n\rightarrow \infty}\log \Omega(n) = -\sigma(\{R^*_i\})^{1/3} n^{2/3}= m(\{x^*_i\})^{1/3}n^{2/3}.
\eea
From the geometry point of view, to maximize the entropy density, we need to extremize the hyperbolic volume of the ideal polyhedra of quiver.
Using the Lagrange multiplier, we obtain that the maximum of the entropy density as
\bea
\sigma(\{R_i\})=-\frac{1}{\pi}\textit{Vol}(N)=-\frac{1}{\pi}\sum_{i=1}^{d} L(\pi R_i),
\eea
happens at 
\bea
\frac{\partial L(R_1)}{\partial R_1}\bigg\rvert_{R_1=R^*_1} = \frac{\partial L(R_2)}{\partial R_2} \bigg\rvert_{R_2=R^*_2}= ...= \frac{\partial L(R_d)}{\partial R_d}\bigg\rvert_{R_d=R^*_d},
\eea
which leads to $R^*_1= R^*_2 = ...= R^*_d=1/d$. The maximizing point of the hyperbolic volume happens at the symmetric point for the ideal polyhedra built on the regular polygon quivers. Thus, the entropy density is maximized for the regular polygon quiver and we have
\bea
\lim_{n\rightarrow \infty}\log \Omega(n)&=&-\sigma(\{R^*_i\})^{1/3} n^{2/3}\nonumber\\
&=&\frac{1}{\pi^{1/3}}\textit{Vol}(M^*)^{1/3} n^{2/3}\nonumber\\
&=& \left(\frac{1}{\pi}\sum_{i} L(\pi R^*_i)\right)^{1/3}\ n^{2/3}\nonumber\\
&=& \left(\frac{d}{\pi}\ L(\frac{\pi}{d})\right)^{1/3}\ n^{2/3}.
\eea

Equivalently, we can study the variational problem in terms of the Mahler measure of NP and the Bloch-Wigner functions. As we observed the regular polygon quiver is the maximizer of the entropy density. The regularity implies that that all the coefficients, the lenghts of the edges, $a_i$ in the NP of the isoradial quiver are equal and thus can be rescaled to one,
\bea
a^*_1=a^*_2= ... = a^*_d=1,
\eea
and in terms of the fugacity factors, 
\bea
x^*_1=x^*_2= ... = x^*_d=0,
\eea
and all the angles between the edges of the polygons are also equal
\bea
\gamma^*_1=\gamma^*_2= ... = \gamma^*_d= \frac{(d-2)}{d}\pi.
\eea
Thus, in terms of the we have
\bea
\lim_{n\rightarrow \infty}\log \Omega(n)&=&-\sigma(\{z_i^*\})^{1/3}\  n^{2/3}\nonumber\\
&=&-\mathcal{L} \left( m(\{x^*_i\})\right)^{1/3}n^{2/3}\nonumber\\
&=&\left(\frac{1}{\pi}\sum_i D(z^*_i)\right)^{1/3} n^{2/3},
\eea
where $z^*_i$ are the solutions of the  variational problem. More explicit computations about this can be obtained in the examples in chapter 4. 
Thus, in the case of isoradial quivers, the extremum of the BPS entropy density of isoradial quiver is obtained from the Mahler measure of the NP with the same coefficients, i.e. the quiver with the same lengths of the edges.
As we obtained in this section, for the isoradial quiver with cyclic polygon quiver, the regularity of the polygon and equality of the R-charges leads to the extremization of the entropy density. However, the quiver of the isoradial toric Calabi-Yau threefolds such as $\C^3$ and $\cC$ are themselves regular cyclic triangle and square. In other words the maximum of the entropy density is the Mahler measure with the normalized coefficients in the NP.

\subsection{BPS Phase Structure}
In order to study the phase structure of the quivers, we need to explore the possibility of a discontinuity in the $n$-th derivative of the entropy density and find the possible critical non-differentiable points and then study the behavior of the entropy density in the vicinity of the critical points. The general picture of the phase structure in dimer model has a physical interpretation in quivers and is manifested in terms of the jump in the derivative of the BPS entropy density.

In the class of isordial quivers including the isoradially embedded quivers, this interpretation and manifestation can be explicitly studied using the analytic expressions and their properties for the entropy density in terms of the dilogarithm functions. In fact, in the previous parts we obtained the entropy density as the hyperbolic volumes given in terms of the Bloch-Wigner and Lobachevsky functions. Having explicit formulas for the entropy density allows us to study the phase structure of the isoradial quiver via the analytic properties of the entropy density.

In this chapter, first we study the critical (non-analytic) points of the entropy density and extract the phase structure of the quiver. Then, we investigate the behavior of the quiver near the critical points by studying the asymptotic expansions of the entropy density near the non-analytic points.

\subsubsection*{Non-analyticity of Entropy Density and Critical Points}
The entropy density or the hyperbolic volume is the order parameter of the isoradial quiver and the indicator of the phase transition in the system.
The entropy density written as a sum over Bloch-Wigner functions is an analytic convex function except at finite number of critical points. In fact, Bloch-Wigner function is analytic in $\mathbb{C} - \{0,1\}$ which at $0,1$ is continuous but non-differentiable. More precisely, from the mathematical point of view, the non-differentiability of the entropy density can be seen from the divergence of the derivative of entropy density in Eqs. \eqref{ent BW} and \eqref{ent lobach}, at critical points in the following equations,
\bea
\frac{d\sigma}{d\xi} = -\sum_i \frac{d L(R_i)}{d R_i}  \frac{d R_i}{d\xi}= \sum_i \log|2\sin \pi R_i|  \frac{d R_i}{d\xi},
\eea

\bea
\frac{d\sigma}{d\xi} = -\sum_i \frac{d D(z_i)}{dz_i}  \frac{dz_i}{d\xi}= \sum_i \frac{\mathrm{i}}{2}(\frac{\log |1-z_i|}{z_i}+\frac{\log |z_i|}{1-z_i}) \frac{dz_i}{d\xi},
\eea
where we observe that the derivative of BPS entropy density is divergent at following equivalent critical points,
\bea
z^c_i= 0, 1, \quad R^c_i=\frac{\alpha^c_i}{\pi}= 0,1, \quad  a^c_i=2\sin \pi R^c_i= 0, \quad x^c_i\rightarrow -\infty. 
\eea
Thus, we observe that the $R^c$-charges, and the zeros of the quiver edge lengths and divergent fugacity factors are the critical points of the isoradial quivers, in different but equivalent representations. On the other hand, from the physics perspective and using the properties of the Legendre transformation in Eqs. \eqref{ent leg dual} and  \eqref{ent der}, we have 
\bea
\frac{\partial \sigma}{\partial R_i}=  \log a_i= x_i,
\eea
which shows that the first derivative of the entropy density is divergent at the critical points.
This is consistent with the above result from the analysis of the dilogarithm functions.

\subsubsection*{Phase Structure and the Amoeba/Toric Diagram}
The phase structure of the quiver can be studied from the perspective of the dimer model; change of the height function in the crystal model, and the geometric approach of the Amoeba \cite{Zah}. However, in this work, we approach this problem from an analytic perspective and via the critical analysis of the entropy density as a function of $R$-charges. As we will see, these two approaches are in fact related and one can obtain the geometric picture out of the analytic approach. In addition to the analytic information such as order of the phase transition, the critical analysis of the entropy density in terms of the Bloch-Wigner function or Lobachevsky function leads to the phase boundaries. 

In the following, we briefly review the analytic and geometric approaches to the phase structure of the quiver and discuss their relations. In the two variable formalism, the solid/liquid phase structure is given by the Amoeba of the NP in which the boundaries of the Amoeba are separating the unbounded complement of the Amoeba from its unbounded component. Respectively, these are the divergent and convergent domains of the free energy given by the Ronkin function. In dimer mode language, the boundaries of the Amoeba separate different phases of the system with different height fluctuations.

In the dimer model representation, the toric diagram is the dual graph of the tropical limit of the Amoeba, i.e. the spine graph of the Amoeba called $(p,q)$-web, and this is the support of the entropy density. The boundaries of the Amoeba are the phase boundaries. Approaching the boundaries of the Amoeba is equivalent to taking the slopes of the height function to the boundary/inside integer points of toric diagram. In the dual picture, considering the toric diagram as the dual $(p,q)$-web, we can focus on the behavior of the entropy density as a convex function of the slopes, i.e. gradient of the height function, over the toric diagram. In this picture, the boundaries of the toric diagram are the boundaries on which the entropy density is non-analytic. 
Furthermore, the entropy density as a convex function has a maximum, obtained on the symmetric point of the toric diagram,  i.e. the intersection point of the edges of the $(p,q)$-web. As the entropy density is a convex function with support on the toric diagram, it is implied that the maximum of the entropy density is obtained at the symmetric point of the toric diagram.
From the point of view of the Amoeba,
there are two types of bounded and unbounded boundaries in the Amoeba, separating the gas from the liquid and the solid from the liquid phases, respectively. From the crystal/dimer model point of view, the phase boundary between the solid and liquid phases is located in the limit of the zero/one slope, in which the facets of the crystal are approached. 
In principle, from the analytic point of view and by studying the entropy density, both types of the boundaries should be reproduced from the critical limits of the entropy density and its Legendre dual. 
\subsubsection*{Phase Structure and Generalized Amoeba/Toric Diagram}
In this part, we briefly discuss a generalization of the above toric geometry picture for the phase structure and the analysis of the entropy density to the multivariate hyperbolic geometry formulation. However, the proper mathematical formulation of the above heuristic picture of the generalized phase structure remains for future studies. 

First, we define the generalized Amoeba as
\bea
\cA_G=\{\ (\log|a_1|, ..., \log|a_d|)\ |\  P(z,w; \{a_i\})=0\ \}.
\eea
The fugacity factors are analogous of the (generalized) Amoeba coordinates and can be obtained from the entropy density, by using the Legendre transformation properties, as
\bea
(x,y)= \left(\frac{\partial \sigma(s,t)}{\partial s}, \frac{\partial\sigma(s,t)}{\partial t}\right), \quad \{x_i\}= \{\frac{\partial \sigma(\{R_i\})}{\partial R_i}\} .
\eea

In a similar way to Amoeba, the boundaries of the generalized Amoeba are separating the divergent and convergent domains for the Mahler measure. In fact, the hyper-boundaries of the generalized Amoeba $\cA_G$ separate the unbounded component of the generalized Amoeba with convergent Mahler measure of the quiver from the unbounded complement of the generalized Amoeba with divergent generalized Mahler measure. Similar to the Amoeba, the tropical limit of the generalized Amoeba is the spine hyper-graph. 
Analogously, in the hyperbolic representation, we define the generalized toric diagram as the dual to the spine hyper-graph of the generalized Amoeba and the phase boundaries are the hyper-boundaries of the generalized Amoeba and the integer points of the generalized toric diagram. In this generalization, the slopes $(s,t)$ of the dimer model is generalized to the $R$-charges $(R_1, R_2, ..., R_d)$ and they are defined as the derivative of the Ronkin function $N(x,y)$ and the generalized Mahler measure $m(\{x_i\})$, respectively,
\bea
(s,t)= \left(\frac{\partial N}{\partial x}, \frac{\partial N}{\partial y}\right), \quad \{R_i\}= \{\frac{\partial m}{\partial x_i}\}.
\eea
Similar to the dimer model representation in which the Amoeba is the phase structure and the support of the free energy, and the toric diagram is the support of the entropy density, there is a generalization of the geometric picture for the phase structure of the isoradial quiver. In the hyperbolic representation, the generalized Amoeba is the phase structure and the support of the free energy of isoradial quivers. The generalized toric diagram is the support of the entropy density on which the entropy density is an analytic convex function and on the boundaries of the generalized toric diagram, it is non-analytic. The boundary integer points of the generalized toric diagram are the critical hyper-surfaces. Similar to the two-variable case, the symmetric point of the generalized toric diagram is the maximizing point of the entropy density.

\subsubsection*{Analysis of Entropy Density Near the Critical Points}
Having obtained the critical points, the next step is to analyse the entropy density near the critical points.
We study the asymptotic expansion and analytic properties of the BPS entropy density in the critical limits. The phase structure of the BPS sector of the isoradial quiver including the boundaries and the order of the phase transition, can be extracted from the analytic behavior of the BPS entropy density near the critical points.

The critical behavior of the entropy density is in the zero limit and boundary point of the generalized toric diagram and equivalently is obtained by approaching the boundary of the generalized Amoeba from the convergent domain. In terms of the shape parameters, the expansion of the entropy density near the critical point $z_i\rightarrow 0$ can be obtained from the definition of the Bloch-Wigner function, and at the critical point $z_i\rightarrow 1$ from the property $D(z)= - D(1-z)$, as
\bea
\lim_{z_i\rightarrow 0,1} \sigma\left(\{z_i\}\right) \approx \frac{1}{\pi}\sum_i \pm |z_i| \log |z_i|,
\eea
where plus (minus) sign is for $z_i= 1 (0)$ critical points.

Similarly in terms of the dihedral angels and R-charges, at critical points $R_i\rightarrow 0, 1$, the entropy density can be expanded in terms of the $R$-charges. First, notice that the the Lobachevsky function satisfies
\bea
\frac{d^2 L(\alpha)}{d\alpha^2} = -\cot\alpha.
\eea
Then, by integrating the series expansion of the cotangent function, and using Eq. \eqref{ent lobach}, one obtains the asymptotic expansion of the entropy density,
\bea
\sigma(\{R_i\}) = -\sum_{i=1}^d \big(R_i-R_i\log|2\pi R_i|+\sum_{n=1}^{\infty}\frac{\zeta(2n)}{n(2n+1)}R_i^{2n+1}\big),
\eea
which is convergent for $|R_i|\leq 1$.

In the limit $R_i \rightarrow 0,1$, entropy density becomes
\bea
\lim_{R_i\rightarrow 0,1}\sigma(\{R_i\}) = \sum_{i=1}^d R_i\log R_i + O(\pi R_i).
\eea
In other words, near the origin of the generalized toric diagram, we obtain the Shannon-like entropy in terms of the $R$-charges,
\bea
\lim_{R_i\rightarrow 0,1}\sigma(\{R_i\}) = \sum_{i=1}^{d-1} R_i \log R_i- \left(\sum_{i=1}^{d-1} R_i\right)\log \left(\sum_{i=1}^{d-1} R_i\right).
\eea
The above Shannon-like entropy can be seen as a generalization of the Shannon-like entropy in the hexagonal dimer model that has been observed in \cite{Ke-Ok-Sh} and similarly for clover quiver, by approximation of the integral definition of the Lobachevsky function in \cite{Zah}. Furthermore, from the above expansion of the entropy density near the critical points one can also observe that the derivative of the entropy density is not continuous and in fact divergent.

Moreover, in the context of critical phenomena, we expect a notion of universality in the vicinity of the critical points. In our work, we rigorously observe a notion of universality as a general Shannon-like formula for the BPS entropy density in the class of isoradial quivers which leads to a universal phase structure in this class.

\subsection*{Relation to Two-variable Formulation}
\label{reduction}
In this part, we discuss the relations between the hyperbolic formalism to the dimer model formalism of the asymptotics of the isoradial quivers.
The hyperbolic parameterization is based on the the assignment of the statistical weights proportional to the lengths of the edges of the quiver. The weights can be transformed to the minimal number of parameters, i.e. the coordinates of the Amoeba and the K\"{a}hler parameters, as in the two-variable dimer model formulation of the asymptotics. On the other hand, from the geometric point of view of the hyperbolic formalism, a linear combination of the dihedral angles can be seen as the slopes in the dimer model, as we observed in some examples in \cite{Zah}. From the Mahler measure perspective, we can mathematically explain this transformation. The Mahler measure of the quiver, which is the generalization of the generalized Mahler measure on the torus, can be transformed to the Ronkin function.
In fact, the generalized Mahler measure on torus is equal to the Mahler measure of the NP, $P(z,w)= \sum_{(n,m)} \hat{a}_{(n,m)} z^n w^m$ with the choice $\hat{a}_{(n,m)}= e^{-t_{(n,m)}} a^n b^m$,
\bea
m_{(a,b)}(P)= m(\{\hat{a}_{(n,m)}\}).
\eea
In the iosradial quiver, using Eqs. \eqref{Ron mah} and \eqref{Mah iso 1}, the Ronkin function is obtained from the Mahler measure,
as
\bea
\pi N (x,y)= \sum_{(n,m)} \left(L(\hat{\alpha}_{(n,m)})+ \hat{\alpha}_{(n,m)} (-t_{(n,m)}+ nx+ my)\right),
\eea
where $\hat{\alpha}_{(n,m)}=1/2 \arcsin \hat{a}_{(n,m)}$.
Then, comparing with the Legendre transform of the Ronkin function in Eq. \eqref{leg trans} we find that $\hat{\alpha}_{(n,m)}$ are related to the slopes $(s,t)$, as
\bea
\label{change var 2}
s=\sum_{(n,m)} n\ \frac{\hat{\alpha}_{(n,m)}}{\pi},\quad
t=\sum_{(n,m)} m\ \frac{\hat{\alpha}_{(n,m)}}{\pi}.
\eea
The Mahler measure of the normalized NP with $\{a_i=1\}$ is obtained from the Ronkin function by putting $x=y=t_{(n,m)}=0$.

Similarly, for the BPS entropy density we have 
\bea
\sigma(s,t)&=&\cL(m_{(e^x,e^y)}(P))
= -m_{(e^x,e^y)}(P) + s x+ ty\nonumber\\
&=&\cL(m(\{\hat{a}_{(n,m)}\}))= -m(\{\hat{a}_{(n,m)}\}) + \sum_{(n,m)} \frac{\hat{\alpha}_{(n,m)}}{\pi} \log \hat{a}_{(n,m)}.
\eea

As the Mahler measure with all $a_i=1$, reduces to the generalized Mahler measure with $a=b=1$, thus, in terms of the generalized Mahler measure on arbitrary torus, we reproduce the BPS growth rate,
\bea
\lim_{n\rightarrow \infty}\log\Omega(n)=-\mathcal{L} (m_{(1,1)}(P))^{1/3} n^{2/3}=m_{(1,1)} (P)^{1/3} n^{2/3}.
\eea

The phase structure of the isoradial quiver, the generalized Amoeba and its dual, the generalized toric diagram are trivially reduced to the two-dimensional one, by the transformation and change of variables introduced in this section.

\section{Examples}
\label{sec:examples}
In this section, we consider basic concrete examples of isoradial quivers and by using the hyperbolic geometry we compute the free energy, the BPS entropy density and the BPS growth rate in terms of the $R$-charges. Based on these results, we find the BPS Shannon entropy, critical points and the phase structure in these examples.

\subsection{Clover and Conifold Quiver}
The first and simplest examples of the isoradial toric quivers are $\mathbb{C}^3$ and $\cC$ quiver. In this part, we review the results obtained in \cite{Zah} and present them in term of $R$-charges, with no derivation. These results are the special cases of the the general results that will be obtained in section \ref{k-polygon}. 

\subsection*{Clover Quiver\ $\mathbb{C}^3$}
\label{sec:clov}
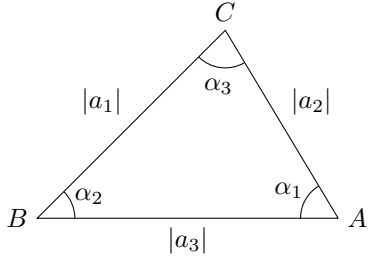
\begin{figure}
\centering
\begin{tikzpicture}[scale=1.25]%,cap=round,>=latex]
\coordinate [label=left:$B$] (B) at (-1cm,-1.cm);
\coordinate [label=right:$A$] (A) at (2.2cm,-1.0cm);
\coordinate [label=above:$C$] (C) at (1cm,1.0cm);
\draw (A) -- node[below] {$|a_3|$} (B) -- node[above left] {$|a_1|$} (C) -- node[above right] {$|a_2|$} (A);
%\pic [draw, -, "$\alpha_0$", angle eccentricity=1.5] {angle = B--A--C};
\pic[draw, -,"$\alpha_1$",angle eccentricity=1.5] {angle=C--A--B};
\pic[draw, -,"$\alpha_2$",angle eccentricity=1.5] {angle=A--B--C};
\pic[draw, -,"$\alpha_3$",angle eccentricity=1.5] {angle=B--C--A};
\end{tikzpicture}
\caption{Triangle for the clover quiver}
\label{triangle}
\end{figure}

%(explain the logic in words)
The clover quiver is related to the NP of $\mathbb{C}^3$ Calabi-Yau, $P(z,w)=1+z+w$. The clover quiver polygon is a triangle with edge lengths $|a_1|$ $|a_2|$, $|a_3|$, in Fig.~\ref{triangle}.
The Mahler measure of NP of the clover quiver, $P(z,w)=a_1z+ a_2w + a_3$, is obtained in \cite{mail}, and its relation to the volume of the hyperbolic ideal tehtrahedron $T$ built on the triangle, are the key to understand the thermodynamics of the clover quiver.

\textbf{Theorem.}  (Maillot \cite{mail}) For $a_1,a_2,a_3 \in \mathbb{C}^3$,
\bea
\pi m (a_1,a_2,a_3)=
\begin{cases}
D\left(\frac{|a_3|}{|a_1|}e^{\mathrm{i} \alpha_2}\right)+\alpha_1 \log |a_1|+ \alpha_2 \log|a_2|+ \alpha_3 \log |a_3| &\text{$\ \{\log |a_i|\}\in \cA$} \\
 \max \pi(\log|a_1|,\log|a_2|,\log|a_3|) &\text{$\ \{\log |a_i|\}\notin \cA$}
\end{cases}.
\eea
It will be instructive for the sake of our discussion to notice that the phase factor $e^{\mathrm{i}\alpha_2}$ is in fact the solution of $P(z,w)=0$ and $|z|=|w|=1$ equations.

The Legendre transformation of the Mahler measure gives the entropy density, as the volume of the ideal tetrahedron built over the triangle $ABC$ and in terms of the $R$-charges we have,
\bea
\label{ent den C}
\sigma_{\mathbb{C}^3}(R_1, R_2, R_3)=-\frac{1}{\pi}D\left(\frac{\sin \pi R_3}{\sin \pi R_1}e^{\mathrm{i}\pi R_2}\right).
\eea
In terms of the Lobachevsky function, the entropy density can be written as,
\bea
\label{ent lob1}
\sigma_{\mathbb{C}^3}(R_1, R_2)=-\frac{1}{\pi} \textit{Vol}\ (T)= -\frac{1}{\pi}\Big(L(\pi R_1) +  L(\pi R_2) +  L\left(\pi(1-R_1-R_2)\right)\Big).
\eea
The Shannon entropy of the clover quiver is obtained in the small $R$-charge limit as,
\bea
\sigma_{\C^3}(R_1,R_2)\approx R_1\log R_1 +R_2\log R_2 - (R_1+R_2)\log (R_1+R_2).
\eea
The BPS growth rate is computed from the maximization of the entropy density, which happens at origin of the Mahler measure,
\bea
\label{clov mahler}
\sigma_{\mathbb{C}^3}(\{R^*_i\})=-m_{\C^3}(\{x^*_i\})=-\frac{1}{\pi}D(e^{\mathrm{i}\pi/3})=-\frac{3}{\pi}
L(\pi/3)\approx 0.32,
\eea
where $R^*_i=1/3$.
Thus, the BPS growth rate can be computed explicitly,
\bea
\label{clov growth}
\lim_{n\rightarrow \infty}\log{\Omega_{\mathbb{C}^3}(n)}\sim \ m_{\mathbb{C}^3}^{1/3} \ n^{2/3}=\frac{1}{\pi^{1/3}} \textit{Vol}\ (T^*)^{1/3}n^{2/3}\approx \ 0.68\ n^{2/3},
\eea
where $T^*$ is the tetrahedron built on the regular triangle.
This result is consistent with the classic result for the asymptotics of the plane partitions, $\lim_{n\rightarrow \infty} \log PP(n)\sim (\zeta(3)/4)^{1/3}\ n^{2/3}$.

\subsection*{Conifold Quiver \ $\mathcal{C}$}
The conifold quiver is related to the NP of $\cC$ Calabi-Yau, $P(z,w)= z w + z + w -1$. 
The thermodynamics of the conifold quiver is encoded in following result for the Mahler measure of the conifold quiver and its relation to the volume of the hyperbolic ideal hexahedron $H$ built on the quiver quadrilateral. 

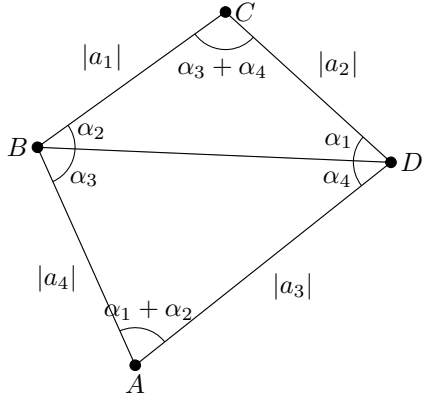
\begin{figure}
\begin{center}
\begin{tikzpicture}
\coordinate (B) at (-2.5,2.2);
\coordinate (A) at (-1.2,-.7);
\coordinate (D) at (2.2,2);
\coordinate (C) at (0,4);
\filldraw
(D) node[align=left, below] {}
-- (B) node[align=center, below, right] {}
(A) circle (2pt) node[align=left, below] {$A$}
-- node[below right] {$|a_3|$} (D) circle (2pt) node[align=center, below, right] {$D$}
-- node[above right] {$|a_2|$} (C) circle (2pt) node[align=left, above, right] {$C$}
-- node[above left] {$|a_1|$} (B) circle (2pt) node[align=left, above, left] {$B$}
-- node[below left] {$|a_4|$} (A);
\pic [draw, -, "$\alpha_1$", angle eccentricity=1.5] {angle = C--D--B};
\pic [draw, -, "$\alpha_2$", angle eccentricity=1.5] {angle = D--B--C};
\pic [draw, -, "$\alpha_1+\alpha_2$", angle eccentricity=1.5] {angle = D--A--B};
\pic [draw, -, "$\alpha_3$", angle eccentricity=1.5] {angle = A--B--D};
\pic [draw, -, "$\alpha_3+\alpha_4$", angle eccentricity=1.5] {angle = B--C--D};
\pic [draw, -, "$\alpha_4$", angle eccentricity=1.5] {angle = B--D--A};
\end{tikzpicture}
\caption{Cyclic quadrilateral for resolved conifold}
\label{fig:quad}
\end{center}
\end{figure}

\textbf{Theorem.} (Vandervelde \cite{Van}) Consider NP of conifold quiver polygon, $P(z,w) = a_1zw+ a_2z+a_3w+a_4$ with $a_1,a_2,a_3,a_4 \in \mathbb{R}$. Suppose there is a non-degenerate convex cyclic quadrilateral $ABCD$ with edge lengths $|a_1|$, $|a_2|$, $|a_3|$, $|a_4|$ in cyclic order, inscribed in a circle and dihedral angles $\alpha_1$, $\alpha_2$, $\alpha_4$, $\alpha_3$, in Fig.~\ref{fig:quad}. The Mahler measure is given by
\bea
\pi m (a_1,a_2,a_3,a_4)=
\begin{cases}
D(\frac{a_3}{a_4}u) - D(\frac{a_1 }{a_2}u) + \alpha_1 \log |a_1|+ \alpha_2 \log |a_2|+\alpha_3 \log |a_3|+\alpha_4 \log |a_4| &\text{$\ \{\log |a_i|\}\in \cA$} \\
\max \pi(\log|a_1|,\log|a_2|,\log|a_3|,\log|a_4|) &\text{$\ \{\log |a_i|\}\notin \cA$}
\end{cases},
\eea
where $u$ is the solution for $w$ of the following equations,
\bea
\label{u}
P(z,w)=0,\quad |z|=1,\ |w|=1.
\eea
In \cite{Zah}, we made the observation that
$u=e^{\mathrm{i} (\alpha_1+\alpha_2)}$.
Then, we obtain the entropy density of the conifold quiver, as the volume of the ideal hexahedron built over the quadrilateral $ABCD$, and in terms of the $R$-charges we have,
\bea
\label{conif ent}
\sigma_{\cC}=-\frac{1}{\pi}  \left(  D\left(\frac{\sin\pi R_3}{\sin \pi R_4}e^{\mathrm{i}\pi (1-R_3-R_4)}\right) - D\left(\frac{\sin\pi R_1 }{\sin\pi R_2}e^{\mathrm{i}\pi (R_1+R_2)}\right)\right).
\eea
The entropy density in terms of the Lobachevsky function the is,
\bea
\label{ent loba C}
\sigma_{\cC}(R_1, R_2, R_3)=-\frac{1}{\pi} \textit{Vol}\ (H)= -\frac{1}{\pi}\Big(L(\pi R_1) +  L(\pi R_2) + L(\pi R_3)+  L\left(\pi(1-R_1-R_2- R_3)\right)\Big).
\eea
In the small $R$-charge limit, we obtain the Shannon entropy of the conifold,
\bea
\sigma_{\cC}(R_1,R_2,R_3)\approx R_1\log  R_1 +R_2\log R_2 + R_3\log  R_3 - (R_1+R_2+R_3)\log (R_1+R_2+R_3).
\eea
We can obtain the Mahler measure of the conifold, 
\bea
m_\cC= m_{\cC}(\{a^*_i\})=-\sigma_{\cC}(\{R^*_i\}),
\eea
where $R^*_i=1/4$. More explicitly,
\bea
m_{\cC}= \frac{1}{\pi} \Big( D(\mathrm{i}) - D(-\mathrm{i})\Big)
=\frac{2}{\pi} D(e^{ \mathrm{i} \pi/2})= \frac{4}{\pi} L(\pi/2)=\frac{2}{\pi} \Im [Li_2(e^{ \mathrm{i} \pi/2})].
\eea
Finally, using $m_{\cC}=\frac{2G}{\pi}\approx 0.58$, we obtain the BPS growth rate of conifold,
\bea
\lim_{n\rightarrow \infty}\log{\Omega_{\cC}(n)}\sim \ m_{\cC}^{1/3} \ n^{2/3}=\frac{1}{\pi^{1/3}} \textit{Vol}\ (H^*)^{1/3}n^{2/3}= \ (\frac{2 G}{\pi})^{1/3}\ n^{2/3}\approx 0.83 \ n^{2/3},
\eea
where $H^*$ is the hexahedron built on the cyclic regular quadrilateral (square).

\subsection{Cyclic $k$-Polygon Quiver}
\label{k-polygon}
In this section, we consider a cyclic $k$-polygon quiver with edges $(a_1, a_2, ..., a_k)$, and compute the thermodynamical observables of the quiver. The results of this section can be seen as the generalization of the previous examples, $\C^3$ and $\cC$ quivers. Similar results for the Mahler measure of the admissible polygons are obtained in \cite{Lal-hyper}.

As a result of Eqs. \eqref{Mah iso 1} and \eqref{Mah iso 2}, the Mahler measure of $k$-polygon quiver is the Legendre dual of the volume of the ideal polyhedron built over the $k$-polygon. Thus, in terms of the $R$-charges we have, 
\bea
m_k(\{a_i\})=
\begin{cases}
-\frac{1}{\pi}\sigma_k(\{R_i\})+ \sum_{i=1}^k R_i \log a_i \quad\quad &\text{$\ \{\log |a_i|\}\in \cA$} \\
\max (\{\log |a_i|\})\quad &\text{$\ \{\log |a_i|\}\notin \cA$}
\end{cases},
\eea
where $a_i= \sin\pi R_i$. The entropy density of the $k$-polygon quiver can be computed as the hyperbolic volume and in terms of the Lobachevsky function,
\bea
\label{k ent lob}
\sigma_k(\{R_i\})=-\frac{1}{\pi} \textit{Vol}\ (N_k)=-\frac{1}{\pi} \sum_{i=1}^k L(\pi R_i).
\eea
The Shannon entropy of the $k$-polygon quiver is obtained from the entropy density in the limit $R_i\rightarrow 0,1$, as,
\bea
\sigma_k(\{R_i\})\approx\sum_{i=1}^{k-1} R_i \log R_i- \left(\sum_{i=1}^{k-1} R_i\right)\log \left(\sum_{i=1}^{k-1} R_i\right).
\eea

As the generalization of the Maillot and Vandervelde resutls in previous examples, our main goal of this part is to compute the entropy density, as the hyperbolic volume given by $\sigma(\{z_i\})=\sum_{i\in T} D(z_i)$, more explicitly in terms of the lengths of the edges of the $k$-polygon quiver and associated dihedral angles. The first step in this generalization is to make two observations from the two examples in previous part.
First, from the results of Maillot \cite{mail}, we observe that for each triangle in the triangulation, there is a $D(z_i)$ term with the shape parameter,
\bea
z_i= \frac{a^1_i}{a^2_i} e^{\mathrm{i}\theta_{a^1_i,a^2_i}},
\eea
where $a^1_i$ and $a^2_i$ are any two of the three edges of the $i$-th triangle which appear in the ratio in a clockwise order, and $\theta_{a^1_i, a^2_i}$ is the angle between the two edges. Furthermore, $\theta_{a^1_i, a^2_i}= \pi-(\alpha^1_i+\alpha^2_i)$, where $\alpha^1_i$ and $\alpha^2_i$ are the dihedral angles associated with $a^1_i$ and $a^2_i$. Notice that the $\alpha^1_i$, $\alpha^2_i$, and $a^1_i$, $a^2_i$  are the dihedral angles and edges of the triangle, including the original edges of the quiver and inside edges of the polygon quiver that are made from the triangulation of the polygon.
For a given polygon with specific edge lengths, there are two ways to obtain the angles between the edges. In fact, there is a formula from trigonometry of the cyclic polygons to determine the angles from the edge lengths. There is also an algebraic method for this problem. As it is obtained in Maillot and Vandervelde, it is natural to generalize these results to conjecture that the angles in the formula are determined from the solutions of the following equations,
\bea
P(z,w)=0, \quad |z|=|w|=1.
\eea
Depending on the polygon, these equations has different number of solutions which is consistent with the geometric picture that for a given triangulation of the polygon there are possibly more that one opposite angles. However, the rigorous analysis of this fact, remains for the future studies.

Second, using the Vandervelde result \cite{Van}, since the opposite triangles in the triangulation are rotated by $\pi$ angle, and the opposite angles are related by $\theta'=\pi-\theta$, we obtain that the opposite triangles in the triangulation have $D$-terms with opposite signs and the angle $\theta$ in shape parameter is rotated by $\pi$ to $\pi-\theta$, thus the contribution of any two opposite triangles in the hyperbolic volume becomes
\bea
D(z_i)-D(z_j)= D\left(\frac{a^1_i}{a^2_i}e^{\mathrm{i}\theta_{a^1_i,a^2_i}}\right) - D\left(\frac{a^1_j }{a^2_j}e^{\mathrm{i}(\pi-\theta_{a^1_j,a^2_j})}\right),
\eea
where $z_i$ and $z_j$ are the two shape parameters of the two opposite triangles and $a^1_i, a^2_i, a^1_j, a^2_j$ are the edges of the triangles that are not shared between the two triangles.

Putting these two results and observations together, now we are able to compute the entropy density of a general $k$-polygon quiver, for a given triangulation. Notice that the $k$-polygon quiver has $k-2$ triangles. Now, we can write the entropy density as the sum over the triangles indexed by $j$ in the triangulation $T$, as 
\bea
\label{entropy-polygon}
\sigma_k(\{\alpha^1_j, \alpha^2_j \})=-\frac{1}{\pi}\sum_{j\in T} \epsilon_j D\left(\frac{\sin \pi \alpha^1_j}{\sin \pi \alpha^2_j}\exp{\mathrm{i} f(\alpha^1_j, \alpha^2_j)}\right),
\eea
where we used the relations $\sin\pi\alpha^1_j=a^1_j$ and $\sin\pi\alpha^2_j=a^2_j$ for the two edges of the $j$-th triangle with their associated deihedral angles $\alpha^1_j$ and $\alpha^2_j$,  and $f(\alpha^1_j, \alpha^2_j)=\pi-\alpha^1_j- \alpha^2_j$ for $\epsilon_j=1$ and $f(\alpha^1_j, \alpha^2_j)=\alpha^1_j+\alpha^2_j$ for $\epsilon_j=-1$.

Next, we explain our particular choice of triangulation, based on which we will obtain the entropy density as the sum over the edges of the quiver. 
For any given $k$-polygon quiver with $k$ even, we triangulate the polygon by connecting the alternate (nearest) vertices of the polygon. Each triangle is made of two consecutive edges of the polygon and the third side is the edge drawn between two alternate vertices and the angle, $\theta$ in the shape parameter, is the angle  between the two edges of the polygon quiver and in front of the third side of the triangle which is an internal edge. In this way, any side of the polygon belongs to one and only one triangle. We continue this triangulation for any inside polygons, made out of the outer triangulation, until the polygon quiver is fully triangulated. In this way, each edges of the polygon is an edge of one and only one triangle and all the edges of the polygon is used once in the triangulation. The inside polygons of the $k$-polygon quiver are triangulated exactly the same way. Notice that for large enough $k$, the inside polygon of the of $k$-polygon is a $k/2$-polygon and similarly $k/4$-polygon inside the $k/2$-polygon and so on. The dihedral angles associated with each edge of any inside polygon are the sum of the two dihedral angles of its outside polygon, associated with the two edges of the outside polygon that are making the triangle with the edge of the inside polygon. 

Having explained our particular choice of triangulation, we can write the entropy density of the $k$-polygon quiver as the sum over the edges of the polygon quiver and the inside polygons,
\bea
\label{entropy-polygon}
\sigma_k(\{a_i, \tilde{a}_i,  \alpha_i, \tilde{\alpha}_i\})=-\frac{1}{\pi}\sum_j \epsilon_j D\left(\frac{a_j}{a_{j+1}}\exp{\mathrm{i}f(\alpha_j, \alpha_{j+1})}\right) -\frac{1}{\pi}\sum_{j'} \epsilon_{j'} D\left(\frac{\tilde{a}_{j'}}{\tilde{a}_{j'+1}}\exp{\mathrm{i}f(\tilde{\alpha}_{j'}, \tilde{\alpha}_{j'+1})}\right),
\eea
where $j$ runs over the alternating edges $a_j$ of the original $k$-polygon quiver and $j'$ runs over the alternating inside edges $\tilde{a}_{j'}$, with dihedral angles $\tilde{\alpha}_{j'}$ of the inside polygons that are made out of the triangulation, and $f(\alpha_j, \alpha_{j+1})=\pi-\alpha_j- \alpha_{j+1}$ for $\epsilon_j=1$ and $f(\alpha_j, \alpha_{j+1})=\alpha_j+ \alpha_{j+1}$ for $\epsilon_j=-1$ and similar definition for $f(\tilde{\alpha}_{j'}, \tilde{\alpha}_{j'+1})$.
In terms of the R-charges, the above equation becomes,
\bea
\label{k ent Blo}
\sigma_k(\{R_i, \tilde{R}_i\})=-\frac{1}{\pi}\sum_j \epsilon_j D\left(\frac{\sin \pi R_j}{\sin\pi R_{j+1}}\exp{\mathrm{i}\pi f(R_j, R_{j+1})}\right) -\frac{1}{\pi}\sum_{j'} \epsilon_{j'} D\left(\frac{\sin\pi \tilde{R}_{j'}}{\sin\pi \tilde{R}_{j'+1}}\exp{\mathrm{i}\pi f(\tilde{R}_{j'}, \tilde{R}_{j'+1})}\right),
\eea
where $\tilde{R}_{j'}$ are the artificial $R$-charges of the inside edges of the inside polygons and defined in similar way as original $R$-charges. In fact, as we explained the relations between the dihedral angles of the original polygon quiver and those of inside polygons, the $\tilde{R}$-charges can be expressed in terms of the original $R$-charges as the sum of the $\tilde{R}$-charges of the edges of the outside polygon that are connected by the inside edge, and continue the sum until we reach the edges of the original quiver. Thus, it is possible to express the second sum in Eq. \eqref{k ent Blo} in terms of the original $R$-charges of the $k$-polygon quiver.
Next, we will argue that the above entropy density in Eq. \eqref{k ent Blo} reduces to the entropy density in Eq. \eqref{k ent lob}. In fact, this is the reason for separating the terms in Eq. \eqref{k ent Blo} into original and inside polygon. In the first sum, each $D$-terms can be written as three $L$-terms of two dihedral angles of the $k$-polygon quiver and one angle between the edges of the $k$-polygon quiver. In the second sum, each $D$-term is the sum of three $L$-terms of three angles which are either the opposite of the angle between the edges of the quiver or the opposite of another angle of an inside triangle made out of triangulation of the inside polygons. In anyways, all the $L$-terms of the opposite angles cancel each other and we are left with $L$-terms of the dihedral angles of the $k$-polygon quivers. The cancellation of the $L$-terms of the opposite angles of any two opposite triangles in the triangulation, can be mathematically explained as following, \cite{Zah},
\bea
D(z_i) + D(1+z_i)
&=& L\left(\arg(z_i)\right)+ L(\arg(1-\frac{1}{z_i}))+ L(\arg(\frac{1}{1-z_i}))\nonumber\\
&&+L(\arg(-\frac{1}{z_i}))+ L(\arg(\frac{z_i}{1+z_i}))+L(\arg(1+z_i).
\eea
Now we can use the Lobachevsky properties as an odd and $\pi$-periodic function, to write $L(\arg(z_i))+L(\arg(-\frac{1}{z_i}))=0$ and
\bea
D(z_i) + D(1+z_i)&=& L(\arg(1-\frac{1}{z_i}))+ L(\arg(\frac{1}{1-z_i}))+ L(\arg(\frac{z_i}{1+z_i}))+L(\arg(1+z_i)\nonumber\\
&=& L(\alpha_1) +  L(\alpha_2) + L(\alpha_3) + L(\alpha_4)\nonumber\\ &=& L(\alpha_1) +  L(\alpha_2) + L(\alpha_3) - L(\alpha_1+\alpha_2+\alpha_3).
\eea
In general, for a $k$-polygon quiver, the number of $L$-terms, $\#_L$, the number of $D$-terms, $\#_D$, and the number of common shared edges, $\#_{E_c}$, are related by
\bea
\#_L=k,\ \#_D=\#_L-2=k-2, \quad \#_L= 3\#_D-2 \#_{E_c}.
\eea

Finally, the BPS growth rate is obtained from the volume of the regular $k$-polygon as a result of maximizing the entropy density discussed in section (\ref{BPS Growth Rate}), as
\bea
\lim_{n\rightarrow \infty}\log \Omega_k(n)\sim \frac{1}{\pi^{1/3}} \textit{Vol}\ (N^*_k)^{1/3}n^{2/3}=\frac{1}{\pi^{1/3}} \left(\sum_{i=1}^k L(\pi R^*_i)\right)^{1/3}=\left( k/\pi L(\pi / k)\right)^ {1/3} n^{2/3}.
\eea
 
It is interesting to observe that, as $k\rightarrow \infty$, using $L(\theta)\approx -\theta \log\theta$, we have
\bea
\label{k-polygon index}
\lim_{n,k\rightarrow \infty}\log \Omega_k(n)\sim\left(\log k\right)^ {1/3} n^{2/3}.
\eea
The volume of the ideal (regular) $k$-polygon increases as $k$ increases, thus the BPS entropy density and growth rate is a growing function of $k$. This is consistent with Eq. \eqref{k-polygon index}. 

\subsubsection*{The Phase Structure of the $k$-Polygon Quiver}
The phase structure of the $k$-polygon quiver is similar to the phase structure of the general isoradial quiver. The boundaries of the Amoeba determined by the critical limit $R_i\rightarrow 0,1$ are the phase boundaries of the $k$-polygon quiver. 
%the hypersurfaces that $R_i\rightarrow 0,1$...? 
From the hyperbolic 3-manifold point of view, the critical point $R_i\rightarrow 0,1$ for all the quiver edges $i$, is a singular point of the hyperbolic volume differential form since in this limit all the dihedral angles shrink to zero/$\pi$ and all the quiver edges shrink to zero and thus the polygon and its associated polyhedron collapse to a point. 

The phase structure of $\C$ and $\cC$ is the special case of the phase structure of the $k$-polygon quiver. As we observed in \cite{Zah}, the entropy densities of the clover and conifold quivers are continuous across the boundaries of their Amoeba and their first derivatives are not continuous and in fact are divergent approaching the boundaries of the Amoeba. And thus we observed a possible phase transition in these quivers.

\subsection{Hirzebruch Quiver $\mathbb{F}_0$ at Isoradial Point}
In this work, we mostly studied the isoradial quivers. However, there is another class of quivers which are obtained from the isoradial embedding of the non-isoradial quivers. In this section, we consider one of the simplest examples of this type of quivers, namely the Hirzebruch $\mathbb{F}_0$ quiver characterized by the NP, $P(z,w)=z+w+\frac{1}{z}+\frac{1}{w}-l$, at isoradial point, $l=4$. The parameter $l$ is associated with the inside point of the toric diagram and the size of the bounded component of the complement of the Amoeba, i.e. the hole inside the Amoeba. At the isoradial point, the hole inside the Amoeba which is the gas phase in the phase structure of the quiver shrinks to a point and the quiver becomes isoradial, \cite{Zah}. From the thermodynamics point of view,  the isoradial embedding is assigning the isoradial weights for the isoradial quiver. In the isoradial embedding the contribution of the gas phase to the entropy density and the asymptotics is eliminated. Thus, at isoradial point, we expect that the BPS index is computed via the liquid phase entropy given by the hyperbolic volume. 

The dual of the Hirzebruch quiver at general $l\ne 4$ is the square-octagon dimer model \cite{Kenlec}. Isoradial embedding of the square-octagon dimer is by shrinking the edges of the square part of the graph to zero and reducing the square-octagon graph to a square graph. This process is quite similar to the Seiberg duality discussed in this quiver, \cite{Franco:2005rj}. 

The BPS index of the Hirzebruch quiver at isoradial point is computed in \cite{Zah} and we briefly summarize the results, without the derivation, in the following. The BPS growth rate is obtained from the Mahler measure and the Mahler measure of the Hirzebruch quiver, $m_{\mathbb{F}_0}(l)$ for $l=4$, is computed in \cite{Vil},
\bea
\label{mah P1}
m_{\mathbb{F}_0}(4)=\frac{1}{2} \Bigg(\log \frac{1}{16}-\frac{1}{4}\ {}_4 F_3\left(\begin{matrix}&3/2& &3/2& &1& &1&\\ \hspace{.1cm} &2&
&2& &2&\end{matrix};\ 1\right)\Bigg).
\eea 
Furthermore, we obtained,
\bea
m_{\mathbb{F}_0}(4)=\frac{4 G}{\pi}= 4\pi^{-1} D(e^{\mathrm{i}\pi/2})= 4\pi^{-1} \Im [Li_2( \mathrm{i} \pi/2)]\approx 1.16,
\eea
and we observe $m_{\mathbb{F}_0}(4)=2 m_\cC= \frac{2}{\pi} \textit{Vol} (H^*)$.
Thus, the BPS growth rate in this quiver is
\bea
\log{\Omega_{\mathbb{F}_0}}(4)\sim \ m_{\mathbb{F}_0}(4)^{1/3} \ n^{2/3}= \ (\frac{4G}{\pi} )^{1/3}\ n^{2/3}\approx 1.05 \ n^{2/3}.
\eea
The above result can be explained from the isoradial embedding of the Hirzebruch quiver and the square-octagon dimer model. In fact, by shrinking the square part of the dimer graph to zero we reduce the square-octagon graph to its Seiberg dual, the square graph which is the isoradial quiver, and that is the conifold quiver. There are two topologically distinct ways of doing the reduction by sending either of the two sides of the square to zero, before the other one. In this way, it is natural to obtain the BPS growth rate of the Hirzebruch quiver as twice as that of the conifold quiver. Moreover, the process of isoradial embedding suggests an approach to obtain the Mahler measure and entropy density of the isoradially embedded quiver. First, notice that in the Hirzebruch quiver, there are five coefficients of the NP corresponding to four boundary points and one inside point of the toric diagram. The inside point is fixed at $l=4$ in the isoradial limit and thus we are left with four coefficients which are associated with the four edges of the quiver. In other words, the edges of the isoradial quiver corresponds to the boundary points of the toric diagram. Consider the NP of the Hirzebruch quiver in the isoradial limit, $P(z,w)= az +bw +c/z+d/w -4$. As there are two ways to reduce the quiver, there must be two topologically different assignments of the four coefficients of the NP to the four edges of the quiver and thus we have will have two distinct quadrilaterals. These two options can be seen as the resolved and the deformed conifolds. Then, we can use our result for the $k$-polygon quiver and/or the Vandervelde results \cite{Van}, in the case of quadrilateral, to compute the Mahler measure and entropy density from two distinct ideal polyhedra in this case.
The details of the complete analysis of this combinatorial problem require a separate comprehensive study and thus remains for future research. However, in the following we heuristically discuss the variational problem of the entropy density and the properties of the maximal point of the entropy density of the isoradially embedded quivers which leads to the growth rate. We also study the relations to the phase structure of the quiver in the isoradial limit. 

\subsubsection*{Phase Structure and Variaitial Problem of Entropy Density}
As we discussed, in the isoradial limit, the hole of the Amoeba is shrunk to a singular point, which we call it the gas point. The phase structure of the isoradially embedded quiver is almost explained in a same way as the isoradial quiver, in terms of the Amoeba and toric diagram. However, in this case there is a gas point in the phase structure which will be briefly explained in the following.

In the isoradially embedded quivers, the extremization problem of the entropy density is slightly more complicated than the isoradial quivers. As we explained, the isoradial limit of the non-isoradial quiver can be seen as deforming of the quiver and its Amoeba. For the quiver, the isoradial point in moduli space of the parameters is a unique point that all the genus one K\"{a}hler moduli associated with the holes sizes in the amoeba are taken to the isoradial one, and all the genus zero  K\"{a}hler moduli are taken to one, and thus the quiver polygon becomes regular cyclic polygon. For the Amoeba, in the isoradial limit all the holes of the Amoeba are shrunk to points. 

In terms of the quiver thermodynamics, in the isoradial limit in which the original non-isoradial quiver is taken to a cyclic regular polygon, all of the coefficients $a_i$ are taken to one except the coefficients which are associated with the sizes of the gas phase. These coefficients are taking to the isoradial points which are not necessary one and  they can be different for different coefficients associated with different holes. In general. there are different possibilities to take the isoradial limit depending on the number of the holes and the geometry of the holes.

The geometric symmetries of the system implies that the entropy density is a symmetric function of the slopes on the toric diagram, meaning that there is a symmetric point on the toric diagram which is the intersection point of the edges of the dual graph $(p,q)$-web, and the entropy density behaves symmetrically around this point. Furthermore, the entropy density is a convex function on the toric diagram. As a result, the symmetric point of the toric diagram should be the maximizing point of the entropy density which is also a smooth analytic point, if there is no inside point in the toric diagram. In the Hirzebruch quiver, the zero of the toric diagram is an inside point corresponding to the gas point. As the inside point is the symmetric point of the toric diagram, we conclude that the gas point is the singular maximizing point of the entropy density in the Hirzebruch quiver.
From the thermodyanmics of the dimer model \cite{Ke-Ok-Sh}, we know that the height fluctuations in the gas phase are unbounded and thus it is expected that, in the isoradially embedded quiver, the maximal point of the entropy density to be the gas point which is the internal point of the toric diagram. Thus, in the isoradial limit, the gas point is the maximal point of the entropy density.

\section{Discussion and Future Directions}
The hyperbolic 3-manifolds provide a geometric and physical framework to study the thermodynamics of the BPS sector of the isoradial toric quiver gauge theories. This framework is physically well-motivated via the concrete relation between the dihedral angles and R-charges. In this work, we developed the asymptotic analysis of the isoradial quiver gauge theories and obtained the thermodynamical observables of the BPS sector, such as free energy, entropy density, growth rate and phase structure, in terms of the $R$-charges of the fields of the quiver. 

The relation between the dimer model and hyperbolic geometry is observed in \cite{Ken-sur, Ken-iso}. However, in this work we reproduced these results from the view point of quivers and proceed one step further to the analysis of these results in order to study the asymptotics of the isoradial quiver.
A natural implication of our results would be to shed light on the not deeply understood relation between the dimer model and the hyperbolic 3-manifolds.

In this work, we have considered and studied a variational problem related to the BPS entropy density and we obtained $R^*$-charges which maximizes the BPS entropy density. The connection between this variational problem and the $a$-maximization \cite{in-we} and $Z$-maximization \cite{ma-sp} problems is an interesting problem and remains for future studies.

In this paper, we mostly considered isoradial quivers with a brief explanation of one example of isoradial embeddings of the non-isoradial quivers in the case of Hirzebruch quiver. Our next step in future is to study the isoradial limit of any toric quiver and study the triangulation of the toric diagram and derive the thermodynamical observables of the isoradially embedded toric quivers. Perhaps, beside hyperbolic geometry, the toric geometry data and topological vertex formalism help us to explicitly study the asymptotics. The toric polygons beyond $\mathbb{C}^3$ and $\cC$, have vertices inside the toric diagram and thus should be considered in the isoradial limit. However, as the toy model of these quivers we considered the isoradial $k$-polygons with no gas point inside. Furthermore, in the isoradial class, we can explicitly compute the Ronkin functions and its integral over the amoeba, and thus the computation of the genus-zero Gromov-Witten invariant is computationally feasible.

Moreover, as the isoradial and isoradially embedded quivers have no gas phase, the result of this paper can be interpreted as the liquid phase contribution to the free energy, BPS entropy density and growth rate of any general possibly non-isoradial quiver. A natural continuation of this study is to consider the general (incluidng non-isoradial) quiver and obtain explicit results for the thermodynamical observables. Perhaps, the main difficulty is in the computation of the contribution of the gaseous phase of the non-isoradial quivers in the observables, starting from the simplest example, the Hirzebruch quiver. This contribution is perhaps beyond the hyperbolic geometry techniques that is used in this articles. For example, contribution of the gas phase into the BPS entropy density is beyond the hyperbolic volume of the 3-manifolds.

In this work, we observed the relation between the Schl\"{a}li differential form and the thermodynamic observables of the isoradial quivers. To continue in this line of research, further studies on the statistical mechanics of the isoradial quiver gauge theories and dimer models from the hyperbolic geometry point of view are highly interesting. An immediate suggestion would be the investigations for the possible implications and gauge-theoretic interpretations of the topological invariants of hyperbolic geometry such as Dehn invariant in the context of quiver gauge theories.

A closely related problem is the asymptotic study of the reduced quivers which are associated with the two-dimensional crystal model \cite{Nish}. A possible approach is to formulate the two-dimensional analogue of asymptotic analysis of the three-dimensional crystal melting, using the tropical geometry techniques related to Amoeba and Algae. In fact, the Amoeba as the limit shape and its Legendre dual in the large deviation problem determine the thermodynamics of the reduced quivers. The critical asymptotic analysis of the reduced quivers can be performed using the similar techniques from the hyperbolic geometry. In fact, the crystal model of the reduced quiver is obtained as the 2d slices of the crystal model associated with the original quiver and thus the phase structure of the original and reduced quivers perhaps are closely related via the action of a projection.  

Statistical mechanics of the dimer model, especially in the isoradial class, contains more information about the observables of the quivers guage theory. Further studies on the asymptotics of the superconformal indices of toric isoradial quivers and partition function of supersymmetric gauge theories on 3-spheres, in terms of the partition function of the isoradial dimer model is a promising direction for future studies. Perhaps, the starting point would be a related observation in this context, made in \cite{Ya-YBE}.

\section{Appendix}
\subsubsection*{Dilogarithm, Bloch-Wigner and Lobachevsky Functions}
The dilogarithm function is the generalization of the logarithm function, defined by a power series,
\bea
\textit{Li}_2 (z)= \sum_{n=1}^\infty \frac{z^n}{n^2}, \quad \textit{for}\ |z|<1,
\eea
and it has the analytic continuation,
\bea
\textit{Li}_2 (z)= -\int_0^z \frac{\log(1-t)}{t} dt.
\eea
There are closely related cousines of the dilogarithm function namely the Lobachevsky function and Bloch-Wigner function. In fact,
there is a modification of the dilogarithm function called Bloch-Wigner function,
\bea
D(z)= \Im(Li_2(z))+ \log|z| \arg(1-z), \quad z\in \mathbb{C}\setminus \{0,1\},
\eea
which analytically extends the dilogarithm function to the entire complex plane except at points 0 and 1, where it is continuous but not analytic. In this study, the non-analytic points of the Bloch-Wigner function plays a crucial role as the critical points of the entropy density and specify the phase structure of the quiver. The Bloch-Wigner function satisfies the 6-fold symmetry \cite{Zag},
\bea
D(z)=D(1-\frac{1}{z})=D(\frac{1}{1-z})=-D(\frac{1}{z})=-D(1-z)=-D(\frac{-z}{1-z}).
\eea
There is continuous, $\pi$-periodic function for all $\theta$, called Lobachevsky function $L(\theta)$ defined as \cite{Rat},
\bea
L(\theta)= -\int_0^\theta \log|2 \sin (t)|\ dt.
\eea
For any positive integer $n$, $L(\theta)$ satisfies 
\bea
L(n\theta)=n \sum_{j=0}^{n-1} L(\theta+ \frac{j\pi} {n}).
\eea

The Bloch-Wigner function and the Lobachevsky function are related, \cite{Lew},
\bea
\label{ent clov loba}
D(\zeta)=L(\theta_1)+ L(\theta_2) + L(\theta_3)= L(\theta_1)+ L(\theta_2) - L(\theta_1+\theta_2),
\eea
for $\zeta\in \mathbb{C}\setminus \mathbb{R}$, $\theta_1=\arg(\zeta)$, $\theta_2=\arg(1-\bar{\zeta})$, $\theta_3=\arg(\frac{1}{1-\zeta})$. 
Moreover, we have
\bea
D(e^{\mathrm{i}\theta})=\Im[Li_2(e^{\mathrm{i}\theta})]= 2L(\frac{\theta}{2}).
\eea
\subsubsection*{Hyperbolic Volume Computations and Dilogarithm}
The ideal tethrahedron in $\H$, denoted by $T(z)$, has four vertices lie on $\partial \H$, $0,1,\infty, z$, and characterized by the shape parameter $z$, up to an isometry.
The volume of the hyperbolic tetrahedron with dihedral angles $\alpha_i$, i.e. arc length's of the edges is given by \cite{mil}, 
\bea
\textit{Vol}(T) = \sum_i L(\alpha_i).
\eea
Using the definition of the Lobachevsky function we have
\bea
d\textit{Vol}(T)&=& -\sum_i \log (2\sin\alpha_i)\ d\alpha_i\nonumber\\
&=& \log |z| d\arg(1-z) - \log |1-z| d \arg(z)\nonumber\\
&=& d (\Im{Li_2(z)}) - d (\log|z| \arg (1-z)),
\eea
where in the second line we used the sine law and in the third line we used $d (\Im{Li_2(z)})= - \log|1-z|\ d\arg z - \arg(1-z)\ d\log|z|$.
Using $\eta(z,w)= \log |z| d \arg w- \log |w| d \arg z$, the volume differential form becomes 
\bea
d\textit{Vol}(T)= \eta(z,1-z)= d D(z),
\eea
and thus the volume of the tetrahedra $T(z)$ is, \cite{Ne-Za},
\bea
\textit{Vol}(T(z))= D(z).
\eea
If the polydehron $N$ is traigulated to the tetrahedra $T(z_i)$, we have
\bea
d \textit{Vol} (N)=\sum_i\eta(z_i, 1-z_i)=\sum_i d D(z_i) = d (\sum_i D(z_i)) .
\eea
and thus the volume of the ideal polydehra is sum of the volumes of the ideal tetrahedra $\textit{Vol}(T(z_i))$,
\bea
\textit{Vol}(N)= \sum_{i=1}^n \textit{Vol}(T(z_i))= \sum_{i=1}^n D(z_i).
\eea

\bibliographystyle{plain}
\bibliography{references}

\end{document}